\documentclass[twocolumn,10pt]{asme2ej}
\usepackage{epsfig} 

\usepackage{url}
\usepackage{enumerate}
\usepackage{amssymb,amsmath,mathtools}
\usepackage{graphicx,subfigure}
\usepackage{float}
\usepackage[T1]{fontenc}
\usepackage[font=footnotesize]{caption}
\usepackage{fancyvrb}
\usepackage{color}
\usepackage{afterpage}
\usepackage{slashbox}
\usepackage{setspace}
\usepackage{siunitx}

\usepackage{lineno} 

\title{Verification Studies for the Noh Problem using Non-ideal Equations of State and Finite Strength Shocks}

\author{Sarah C. Burnett
    \affiliation{Department of Mathematics\\
	University of Maryland College Park\\
		Email: burnetts@math.umd.edu \\}	}

\author{Kevin G. Honnell
    \affiliation{Computational Physics Division\\
	Los Alamos National Laboratory\\
        Email: kgh@lanl.gov}}

\author{Scott D. Ramsey\\
       Theoretical Design Division\\
	Los Alamos National Laboratory\\
	Email: ramsey@lanl.gov \\
       {\tensfb Robert L. Singleton Jr.}
    \affiliation{Computational Physics Division\\
	Los Alamos National Laboratory\\
        Email: bobs1@lanl.gov}}

\begin{document}

\maketitle    

\begin{abstract}
{\it

The Noh verification test problem is extended beyond the commonly studied ideal gamma-law gas to more realistic equations of state (EOSs) including the stiff gas, the Noble-Abel gas, and the Carnahan-Starling EOS for hard-sphere fluids. Self-similarity methods are used to solve the Euler compressible flow equations, which in combination with the Rankine-Hugoniot jump conditions provide a tractable general solution. This solution can be applied to fluids with EOSs that meet criterion such as it being a convex function and having a corresponding bulk modulus. For the planar case the solution can be applied to shocks of arbitrary strength, but for cylindrical and spherical geometries it is required that the analysis be restricted to strong shocks. The exact solutions are used to perform a variety of quantitative code verification studies of the Los Alamos National Laboratory Lagrangian hydrocode FLAG.

}
\end{abstract}

\section{Introduction}

The Noh problem \cite{noh, gehmeyr, ramsey} is a well-studied and widely used verification test problem in the field of computational hydrodynamics of compressible fluids.  
Typically, it is applied to an ideal polytropic gas \cite{gehmeyr, Jones:1996} and serves as a tool for testing the accuracy of sophisticated hydrocodes.  Axford first obtained analytic solutions to the Noh problem for the non-ideal stiff gas EOS and for a specialized form of the Mie-Gr\"{u}neisen EOS \cite{axford}.  Recently the topic was revisited and generalized by Ramsey et al.\cite{ramsey}, whose formal solutions we employ here. As the Noh problem is also an example of the broader class of Riemann problems, the one-dimensional planar studies of Menikoff and Plohr, Banks, Whitham, and Toro are also relevant to the development of exact, non-ideal solutions in this context \cite{menikoff1989riemann, banks2010exact, whitham2011linear, toro2013riemann}. \\

In this work we review the general solution of the Noh problem, which is most rigorously derived using Lie group methods \cite{ramsey}, and then apply the general solution to three more realistic, non-ideal EOSs: the stiffened gas \cite{harlow1966formation, harlow, wei, Cole:1948,Courant:1944, Eliezer:2002, weixin} , the Noble-Abel gas \cite{johnston, Corner:1950, baibuz1986covolume, Khaksarfard2010, Chenoweth:1983, emanuel, annamalai, hirn, Partington:1949, cagle, van2004continuity, xiang2005corresponding}, and the Carnahan-Starling EOS for a model fluid composed of hard spheres \cite{carnstar, Hansen:1986, song}.  
The exact solutions for these non-ideal EOSs are then compared to the numerical results obtained using the Los Alamos National Laboratory (LANL) compressible-flow code FLAG \cite{burton90, burton92, burton94, caramana}.\\

\begin{figure}[H]
\centering
\includegraphics[width=3.25in]{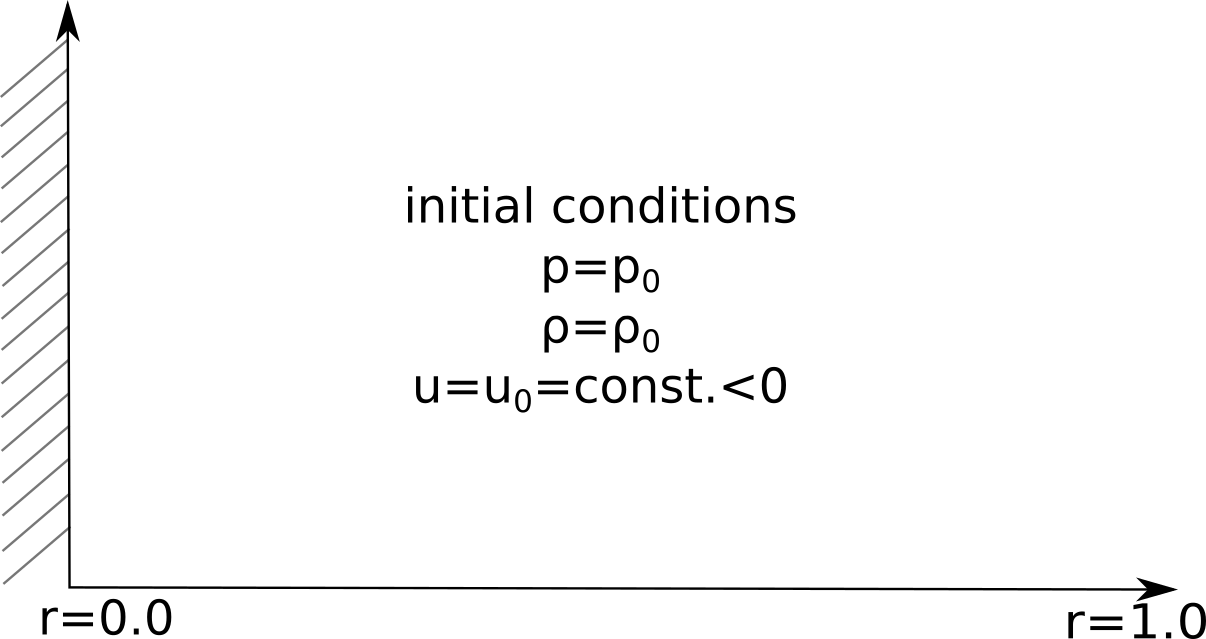}
\caption{The planar Noh problem ($t=0$) -- a uniform inwardly flowing fluid with  pressure $p_0$ and density $\rho_0$ impinges on a hard wall (or an axis or a point in cylindrical or spherical geometries). } \label{fig: IC}
\end{figure}

The Noh problem consists of a strong shock forming as a compressible gas moves at a constant velocity towards a rigid wall (in planar geometry) or as the gas is compressed with constant radial velocity towards a central axis (cylindrical) or a point (spherical) in the curvilinear cases. 
(See Figs. \ref{fig: IC} and \ref{fig: shock}.) 
Alternatively, and equivalently,  in planar geometry the problem can be visualized as shock wave driven by a piston moving at a constant velocity traveling into a quiescent ideal gas \cite{gehmeyr}. 
The problem is defined by specifying the geometry; the EOS of the fluid; and the initial velocity, density, and specific internal energy of the fluid. 
It is typically applied to an ideal gas initialized at zero pressure and energy (a convenient, but rather unrealistic, initial condition corresponding to a temperature of absolute zero), giving rise to an infinitely strong shock since the speed of sound at the initial conditions is zero.  
In this paper, we explore extensions of the Noh problem beyond these idealized conditions. \\

The purpose of this investigation is threefold: to demonstrate the feasibility of extending hydrodynamic verification studies to more physically realistic materials and initial conditions; to use the results to assess a particular hydrocode's ability to accurately capture the shock formation in the Noh problem; and to elucidate the spatial convergence order of the hydrocode under non-ideal gas conditions.
In previous studies, acceptable accuracy has been found to be first order in spatial-convergence tests for hydrocodes including any discontinuities \cite{majda, banks, godunov54, godunov59}.
\\

We begin by reviewing the general one-dimensional solution of the Noh problem derived using the conservation equations, self-similarity methods, and Rankine-Hugoniot jump conditions. 
Following this, we describe the EOSs and the motivation for choosing these particular forms for the verification studies. 
Specific analytic solutions for each EOS are then compared with numerical results obtained using the Lagrangian hydrocode FLAG, so that the spatial-convergence order can be calculated.  
Outcomes for the relative error between numerical and exact results and the observed spatial-convergence order are presented for all three non-ideal EOSs, with a particular focus on planar geometry, where finite strength shocks can also be considered.  Results for infinitely strong shocks in cylindrical and spherical geometries (still one-dimensional) using the Noble-Abel EOS are also presented.  We conclude with a summary of our findings and ideas for future work.

\section{Theory}
An analytic solution to the Noh problem can be constructed from piecewise isentropic solutions of the inviscid one-dimensional Euler compressible flow equations in combination with the Noh initial and boundary conditions and the Rankine-Hugoniot jump conditions \cite{Eliezer:2002} applied across the shock interface. In the absence of an external field (i.e. no gravity or outside forces), the one-dimensional Euler equations for mass, momentum and internal energy are\cite{harlow, ovsiannikov}

\begin{align} 
\frac{\partial \rho}{\partial t} + u \frac{\partial \rho}{\partial r} + \rho \left(\frac{\partial u}{\partial r} + \frac{mu}{r}\right) &= 0  \label{eqn:eulermass1} \\
\frac{\partial u}{\partial t} + u \frac{\partial u}{\partial r} + \frac{1}{\rho} \frac{\partial p}{\partial r} &= 0  \label{eqn:eulermomentum1} \\
\frac{\partial p}{\partial t} + u \frac{\partial p}{\partial r} + K \left(\frac{\partial u}{\partial r}+ \frac{mu}{r} \right) &= 0 \label{eqn:eulerenergy1}
\end{align}

\noindent where  $t$ denotes time and $r$ represents distance from the rigid wall, center axis, or center point depending on whether the geometry is planar, cylindrical, or spherical (distinguished by $m =0$, $1$, or $2$, respectively).  The quantities $\rho$, $u$, $p$, and $K$ represent density, velocity, pressure, and adiabatic bulk modulus, respectively. The density and pressure state variables are related to the specific internal energy, $e$, through an EOS of the form $e(\rho,p)$ or $p(\rho,e)$. The EOS enters into the analysis via the adiabatic bulk modulus, $K$\cite{axford}, defined as 

\begin{equation}
K \left( p, \rho \right) \equiv \rho \left(\frac{\partial p}{\partial \rho} \right)_S = \rho \left( \frac{\partial p}{\partial \rho} \right) _e + \frac{p}{\rho} \left( \frac{\partial p}{\partial e} \right) _\rho \label{eqn:bulk} 
\end{equation}

\noindent where $S$ denotes entropy.  The second identity conveniently expresses the bulk modulus as a function of $e$ and $\rho$, common independent thermodynamics variables in hydrocodes, making it straightforward to derive $K$ from an EOS written in the form $p(e,\rho)$. \\

Lie group methods \cite{ovsiannikov, boyd} leverage the symmetry properties of the one-dimensional Euler equations to construct a similarity variable of the form 
\begin{equation} 
\xi = \frac{ar}{t} \label{eqn:simvar}
\end{equation}
\noindent where $a$ is a constant parameter with units of inverse velocity.  Substitution of Eq. \eqref{eqn:simvar} into Eqs. \eqref{eqn:eulermass1}-\eqref{eqn:eulerenergy1} enables the reduction of the PDEs in question to ODEs in terms of $\xi$,
\begin{align}
- \xi \frac{d \rho}{d \xi} + au \frac{d \rho}{d \xi} + \rho \left(a\frac{d u}{d \xi} + \frac{mau}{\xi}\right) &= 0 \label{eqn:eulermass2} \\
-\xi \frac{d u}{d \xi} + au \frac{d u}{d \xi} + \frac{a}{\rho} \frac{d p}{d \xi} &= 0 \label{eqn:eulermomentum2} \\
- \xi \frac{d p}{d \xi} + au \frac{d p}{d \xi} + K \left(a\frac{d u}{d \xi}+ \frac{mau}{\xi} \right) &= 0. \label{eqn:eulerenergy2}
\end{align}

The Noh solution is piecewise solution, with the discontinuity in the flow interpreted as a shock wave. As a result, we apply the Rankine-Hugoniot jump conditions \cite{Eliezer:2002} at the position of the shock, $r_s(t)$, to ensure conservation of mass, momentum, and energy across the discontinuity 
 \begin{align}
(u_2 - D) \rho_2 &= (u_1-D) \rho_1 \label{eqn:RHmass} \\
p_2 + (u_2-D) \rho_2 u_2 &= p_1 + (u_2 - D) \rho_1 u_1 \label{eqn:RHmomentum} \\
e_2 + \frac{p_2}{\rho_2} + \frac{1}{2} (u_2 - D)^2 &= e_1 + \frac{p_1}{\rho_1} + \frac{1}{2}(u_1 - D)^2 \label{eqn:RHenergy}
\end{align}
where the subscripts $1$ and $2$ represent the unshocked and shocked regions, respectively, and where $D = dr_s/dt$ is the shock velocity. For Eqs. \eqref{eqn:RHmass}-\eqref{eqn:RHenergy} to be invariant under the same group transformations as the PDEs (i.e., for Eqs. \eqref{eqn:RHmass}-\eqref{eqn:RHenergy} to be expressible solely in terms of $\xi$), $D =\text{const.}$ and thus, 
\begin{equation}
 r_s(t) = \frac {t}{a} = D  \cdot t. \label{eqn:shockloc}
\end{equation}  
\begin{figure}[H]
\centering
\includegraphics[width=3.25in]{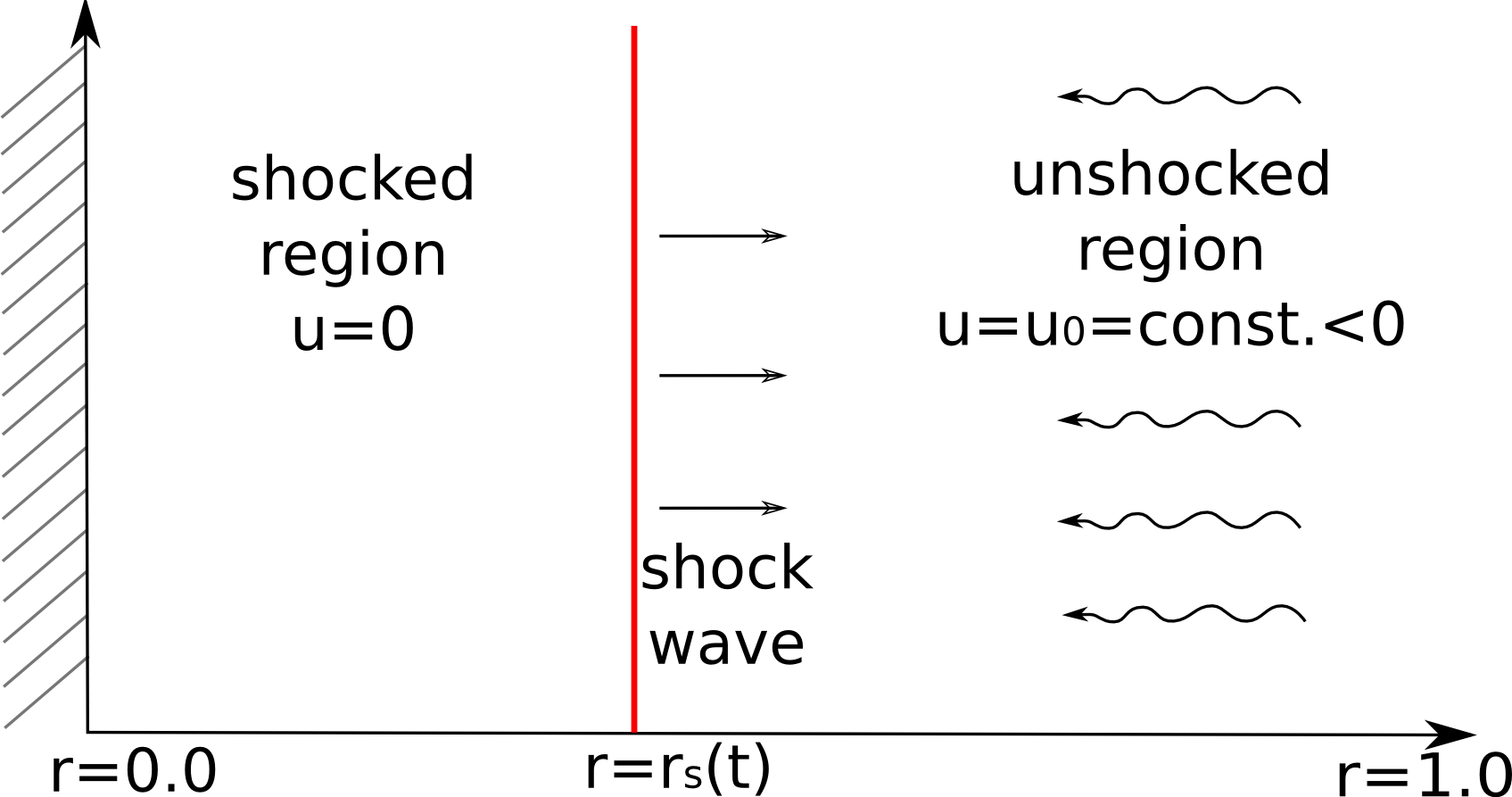}
\caption{The planar Noh problem ($t>0$) --  an outwardly moving shock wave with a stagnant, compressed fluid behind it. } \label{fig: shock}
\end{figure}

The final contribution to the general solution is the conditions of the shocked and unshocked regions characteristic of the Noh problem. Note that $u_1$ and $u_2$ have pre-defined values as consequences of the assumption on the velocity field in the unshocked and shocked regions. The assumed conditions are constant and inwardly directed velocity in the unshocked region, $u_1 = u_0 = \text{const.} < 0$, and zero velocity in the shocked region, $u_2 = 0$.  The effect is the shock wave propagating outward as seen in Fig.~\ref{fig: shock}. 

\subsection{General Noh Solution}
The Noh solution can be found by solving the one-dimensional system of ODEs, Eqs. \eqref{eqn:eulermass2}-\eqref{eqn:eulerenergy2}, and applying the Rankine-Hugoniot jump conditions and the conditions unique to the Noh problem. The following procedure may be used to solve the Noh problem:
\begin{enumerate}
\item By applying the constant velocity condition in the unshocked region, Eqs. \eqref{eqn:eulermass2}-\eqref{eqn:eulerenergy2} can be reduced and Eq. \eqref{eqn:eulermass2} integrated with the initial condition of density to get the general solution for first the unshocked density, then the unshocked pressure from Eq. \eqref{eqn:eulermomentum2}, and lastly, the unshocked energy from Eq. \eqref{eqn:eulerenergy2} by selecting either $K=0$ or $m=0$. 
\item The condition of zero velocity can be applied to Eqs. \eqref{eqn:eulermass2} and \eqref{eqn:eulerenergy2} within the shocked region. From this we notice that it must be the case that both pressure and density in the shocked region are constant. 
\item The density in the shocked region can be calculated by taking the conservation of mass equation from the jump conditions, Eq. \eqref{eqn:RHmass}, and substituting $u_2 = 0$, $u_1=u_0$, $D=1/a$ and $\rho_1$. 
\item The pressure in the shocked region is found by taking the conservation of momentum equation from the jump conditions, Eq. \eqref{eqn:RHmomentum}, substituting for $u_2$, $u_1$, $p_1$ and $\rho_2$ then solving for pressure. 
\item The energy in the shocked region can also be calculated by taking the conservation of energy equation from the jump conditions, Eq. \eqref{eqn:RHenergy}, substituting for $u_2$, $u_1$, $p_1$, $\rho_2$ and $p_2$ and then solving for energy. 
\end{enumerate}
In this fashion, the general solution for the unshocked region is found to be
\begin{align} 
\rho_1 &= \rho_0 \left(1-\frac{u_0t}{r} \right)^m \label{eqn:unshocked} \\
p_1 &= p_0 = \text{constant} \\
u_1 &=u_0 < 0.
\end{align}
\noindent The general solution for the shocked region is
\begin{align} 
\rho_2 &= \rho_0 \left(1-a u_0  \right)^{m+1}  \\
p_2 &= p_0 - \frac{\rho_0 u_0 }{a} \left(1-a u_0 \right)\\
e_2 &= e_1 + \frac{1}{2}u_0^2 - \frac{au_0p_0}{\rho_2}\\
u_2 &= 0.  \label{eqn:shocked}
\end{align} 
Lastly, we can apply the condition $u_2=0$ to the conservation of momentum Rankine-Hugoniot jump condition, Eq. \eqref{eqn:RHmomentum},  and solve for the shock speed to get
\begin{equation}
D = \frac{p_1-p_2}{\rho_2 u_1}. \label{eqn:shockspeed}
\end{equation}
Given Eq. \eqref{eqn:shockloc}, the shock location is determined to be 
\begin{equation}
r_s(t) = D \cdot t = \left(\frac{p_1-p_2}{\rho_2 u_1} \right) t. \label{eqn:shockspeed2}
\end{equation}
The last element required is the EOS in the form $e_2(\rho_2, p_2)$ and $e_1(\rho_1, p_1)$ to create a system of equations with a closed-form solution \cite{ramsey}. If a closed-form solution cannot be found, as with the Carnahan-Starling EOS in Section \ref{CS}, a highly accurate implicit solver can be used. If we relax the definition of verification and allow quasi-analytic solutions to be the exact solution for verification studies, then we can perform verification studies on the Noh problem with any convex EOS.

\subsubsection{Constraints on the Initial Conditions for the Curvilinear cases} \label{constraint}
The constant velocity Noh condition imposes a severe constraint on the initial pressure in cylindrical and spherical geometries. Since $u_1=u_0$ and $p_1=p_0$ are both constant, the energy Euler equation, Eq. \eqref{eqn:eulerenergy2}, reduces to 
\begin{equation}
K \left( \frac{mau}{\xi} \right) =0. \label{eq:condition}
\end{equation}
Since $u$ is a negative constant in the unshocked region and $a$ must be a non-zero constant to have a nontrivial solution then either $m=0$ or $K=0$. In the planar case ($m=0$), this analysis shows that the general solutions can be applied to an EOS that is a convex function and has a corresponding bulk modulus form or with any initial pressure since $K$ need not equal zero. However, if implementing the curvilinear cases ($m \neq 0$), $K$ must be equal to zero by Eq. \eqref{eq:condition}. Since $K = \rho c_s^2$, where $c_s$ is the is the sound speed in the fluid, this implies that $c_s=0$ and that the shock will be infinitely strong. \\

The requirement of a zero initial sound speed in cylindrical and spherical geometries, in turn, constrains the initial pressure. Substituting $K=0$ into Eq. \eqref{eqn:bulk} leads to 
\begin{equation}
\left( \frac{\partial e}{\partial \rho} \right) _p = \frac{p}{\rho^2}
\end{equation}
which can be integrated to give
\begin{equation}
e(\rho, p) + \frac{p}{\rho} = f(p) \label{eqn:arb}
\end{equation}
\noindent where $f(p)$ is an arbitrary function. Within the unshocked region, where $e(\rho,p) = e_1$ and $p=p_1$, Eq. \eqref{eqn:arb} can only be satisfied if either
\begin{equation}
\rho_1 = \text{constant} \label{eqn:case1}
\end{equation}
\begin{center}
or
\end{center}
\begin{equation}
 e_1 + \frac{p_1}{\rho_1} = \text{constant}. \label{eqn:case2}
\end{equation}

Since we have $m=1$ or $2$ and $\rho_1$ is not constant (see Eq. \eqref{eqn:unshocked}), Eq. \eqref{eqn:case1} cannot be true so Eq. \eqref{eqn:case2} must be applied to meet the conditions of conservation and the Noh problem conditions. Substituting the density solution from the unshocked region, Eq. \eqref{eqn:unshocked}, and $p_1 = p_0$, into Eq. \eqref{eqn:case2} provides
\begin{equation}
 e_1 + \frac{p_0}{\rho_0} \left(1-\frac{u_0t}{r} \right)^{-m} = \text{constant}
\end{equation}
implying that $p_0=0$ and $e_1 = \text{constant}$ when $m \neq 0$. Thus, the curvilinear cases can only be studied provided that $p_0 = 0$ \cite{ramsey}. \\ 

The primary focus of this paper is the planar geometry case when $m=0$, where the solution is piecewise constant. An example for the cylindrical and spherical geometries will be demonstrated with the Noble-Abel EOS for infinitely strong shocks. 

\subsection{Equations of State}

In its original incarnation \cite{noh}, and nearly every application since \cite{gehmeyr}, the Noh problem used the ideal polytropic gas EOS \cite{Jones:1996, Courant:1944}, initialized at zero pressure and specific internal energy, to model the thermodynamic behavior of the fluid,

\begin{equation}
p(\rho,e) = \rho e(\gamma -1) \label{eqn:ideal}
\end{equation}
where $\gamma$ is the polytropic constant.  Typically $\gamma$ is set equal to $5/3$, which corresponds to a monatomic ideal gas; however, it can be viewed simply as an adjustable parameter and empirically set to other values depending on the fluid and the thermodynamic path being modeled \cite{Jones:1996}.  \\

While Eq. \eqref{eqn:ideal} admits a particularly simple solution to the Noh problem, its behavior (as well as the conventional initial conditions) is far removed from most materials of practical interest.   Thus there is significant value, from a verification and validation perspective, in extending the domain of applicability of the Noh problem to more realistic EOSs and initial conditions.  We consider here three more complex EOSs:  the stiff gas, useful for metals and liquids;  the Noble-Abel, useful for explosive gases; and the Carnahan-Starling equation for a hypothetical atomistic model fluid composed of rigid, noninterpenetrating, hard spheres. \\

The stiff gas EOS \cite{ramsey, axford, harlow1966formation, harlow, wei, Cole:1948, Courant:1944}, may be viewed as either a modification of the ideal polytropic gas EOS or a simplification (Taylor expansion in density) of the Mie-Gr\"{u}neisen EOS when compressions are modest.  In its most general form, it may be written 
\begin{equation}
p(\rho, e) = p_0 + \left[ c_s^2 - (\gamma -1) \frac{p_0}{\rho_0} \right](\rho - \rho_0) + \rho(\gamma - 1)  (e - e_0) \label{eqn:stiff_gen}
\end{equation}
where $p_0$, $\rho_0$, and $e_0$ denote the pressure, density, and specific internal energy at some convenient reference condition (e.g., the initial conditions of the problem of interest) and  $\gamma$  is an adjustable constant such that $\Gamma = \gamma - 1$ plays the role of an effective Gr\"{u}neisen parameter \cite{Eliezer:2002}. \\

If we choose our reference state such that $p_0$ and $e_0$ are zero, then Eq. \eqref{eqn:stiff_gen} simplifies to the more common form

\begin{equation}
p = \rho e(\gamma -1) +c_s^2 (\rho - \rho_0). \label{eqn:stiff}
\end{equation}

This form of the stiff gas equation has been found to be qualitatively accurate for a variety of metals.  Though similar in appearance to the ideal gas, Eq. \eqref{eqn:ideal}, it enables a condensed-phase material to go into tension as well as compression, and allows use of a physically realistic bulk modulus though the parameter $c_s^2$. Weixin \cite{weixin} has used this form of the stiff gas to model Al, Fe, Cu, and W and reports appropriate values for $c_s$ and $\gamma$.  Harlow and Pracht \cite{harlow1966formation} have used it to model Fe and Al using somewhat different values for $c_s$ and $\gamma$. \\  

Alternatively, if one defines new parameters 
\begin{align}
p_\infty &=\rho_0 c_s^2-\gamma p_0 \\
e_\infty &= \frac{c_s^2}{(\gamma -1)}-e_0-\frac{p_0}{\rho_0}
\end{align}
then Eq. \eqref{eqn:stiff} may be reexpressed as
\begin{equation}
p + p_\infty = \rho (\gamma -1)(e+e_\infty) \label{eqn:stiff_3}
\end{equation}
which is the form used by Courant to model water \cite{Courant:1944} and by Cole to model shock compression of salt water up to 8 GPa \cite{Cole:1948}.  \\

The significance of Eq. \eqref{eqn:stiff_gen} and its simplifications, Eq. \eqref{eqn:stiff} and Eq. \eqref{eqn:stiff_3}, is that the EOS remains simple enough algebraically to permit useful analytical manipulations, such as an exact solution to the Noh problem \cite{ramsey, axford}, but is realistic enough to provide qualitatively accurate predictions for shock compression of liquids and metals, thus enabling code verification studies on physically realistic materials.  In the remainder of this work, we focus on the Eq. \eqref{eqn:stiff} form, where the density, $\rho_0$, is set to the initial density \cite{harlow}. \\

The second EOS we consider is the Noble-Abel equation \cite{johnston, Corner:1950, baibuz1986covolume, Khaksarfard2010, Chenoweth:1983} (also sometimes referred to as the Clausius I \cite{emanuel, annamalai} or Hirn EOS \cite{hirn, Partington:1949, cagle}), which takes the form
\begin{equation}
p(\rho,e) = \frac{\rho e (\gamma -1)}{1-b \rho} \label{eqn:NobleAbel}
\end{equation}
where the adjustable parameter $b$, referred to as the co-volume (units of volume per mass), is intended to account, approximately, for the finite size of the molecules (compare with Eq. \eqref{eqn:ideal}) \cite{annamalai, Partington:1949}.   The original idea for this type of correction to the ideal gas traces its roots to Bernoulli \cite{van2004continuity, xiang2005corresponding}, and reemerges in Hirn's,  Clausius's,  and van der Waals's subsequent EOS developments \cite{clausius, hirn, Partington:1949, van2004continuity} in the nineteenth century. For dilute gases, $b$ may be deduced from the second virial coefficient, and it can take on negative as well as positive values if the temperature is low enough that intermolecular attractive forces decrease the pressure below the ideal gas value.  In hot or strongly compressed fluids, where it is generally more accurate, it takes on positive values (subject to the constraint $\rho b < 1$) and functions to increase the pressure above the ideal gas value.\\

Eq. \eqref{eqn:NobleAbel} has found extensive applicability in ballistics modeling \cite{johnston, Corner:1950, baibuz1986covolume}, where it provides a simple and reasonably accurate EOS for propellant gases at the high densities and temperatures experienced in guns.  It has also been used to model the release of hydrogen into air from high-pressure tanks \cite{Khaksarfard2010}, and in modeling the transfer and discharge of various other compressed gases \cite{Chenoweth:1983}. \\

The third EOS we consider is the Carnahan-Starling equation \cite{carnstar, Hansen:1986, song}.  In contrast to the polytropic ideal gas, stiff gas, and Noble-Abel EOSs, which can be empirically fitted to a variety of fluids, the Carnahan-Starling EOS was developed for a very specific model of a prototypical atomisitic fluid, one composed of noninterpentrating, hard spheres.  The hard-sphere fluid has long been of interest in the field of statistical thermodynamics, in part because the simple form of the interatomic potential greatly reduces calculations of properties such as virial coefficients and radial distribution functions; in part because the specific internal energy is exactly that of a monatomic ideal gas; and in part, because x-ray scattering experiments have shown that the local structure of dense real fluids, including liquid metals, is dominated by the harshly repulsive component of the interatomic potential, which can be well approximated by a hard sphere of an appropriately chosen diameter.  Viewed this way, the hard-sphere fluid becomes a simple reference system about which the effects of interatomic attractions present in real fluids can be added perturbatively \cite{Hansen:1986, mcquarrie}. \\

Carnahan and Starling \cite{carnstar} observed an approximate, serendipitous recursion relationship between the second through the sixth virial coefficients of the hard-sphere fluid which, when postulated to apply equally well to the higher virial coefficients, allowed the infinite series to be summed exactly.  The resulting EOS takes the form
\begin{equation}
p(\rho,e) = \rho e(\gamma - 1)Z(\eta) \label{eqn:CarnStar}
\end{equation}
where $Z(\eta)$ is the compressibility factor
\begin{equation}
Z(\eta) = \frac{1+\eta + \eta^2 - \eta^3}{(1-\eta)^3}
\end{equation}
and $\eta = b \rho$ is the packing fraction of the spheres.  Here again, $b$ denotes the co-volume, now interpreted as the volume per mass occupied by the rigid spheres.  \\

In extensive comparisons with Monte Carlo and molecular dynamic simulations of hard-sphere fluids,  Eq. \eqref{eqn:CarnStar} has been found to be remarkably accurate over nearly the entire fluid range, from the dilute, ideal gas limit, $\eta \to 0$, up to packing fractions, $\eta \simeq 0.45$ \cite{carnstar, Hansen:1986, song}. 

\subsection{Ideal Gas Solution}
The ideal gas EOS, Eq. \eqref{eqn:ideal}, can be rearranged in terms of energy as
\begin{equation}
e = \frac{p}{\rho(\gamma -1)}.
\end{equation}

Ramsey et al. \cite{ramsey} discuss how the general solution to the Noh problem (Eqs. \eqref{eqn:unshocked}-\eqref{eqn:shockspeed2}) can be combined with the EOS to calculate the shock speed $D=1/a$. 
 If we consider only the planar ($m=0$) case then the shock location is
\begin{equation}
r_s(t)=\left[ \frac{A_i+(3-\gamma)  \sqrt{\rho_0} u_0}{4 \sqrt{\rho_0}} \right]t
\end{equation}
where 
\begin{equation}
A_i = \sqrt{16 \gamma  p_0+(\gamma +1)^2 \rho_0 u_0^2}.
\end{equation}
\noindent The density for planar ideal gas solution for the unshocked region is $\rho_1 = \rho_0$ and the density for the planar ideal gas solution for the shocked region is
\begin{equation}
\rho_2 = \rho_0 \left[1 - \frac{4 \sqrt{\rho_0} u_0}{A_i + (3-\gamma ) \sqrt{\rho_0} u_0}\right] .
\end{equation}
As shown in Section \ref{constraint}, for $m=1$ or $2$ solutions exists only in the infinitely strong shock case ($K_0 = 0$, $p_0 =0$) \cite{ramsey}. The shock location is
\begin{equation}
r_s(t) = Dt = - \frac{(\gamma -1)u_0t}{2}.
\end{equation}
\noindent The density in the unshocked region is
\begin{equation}
\rho_1 = \rho_0 \left( 1- \frac{u_0 t}{r} \right)^m
\end{equation}
\noindent and the density in the shocked region is
\begin{equation}
\rho_2 = \rho_0 \left(\frac{\gamma+1}{\gamma-1} \right)^{m+1}.
\end{equation}

\subsection{Stiff Gas Solution}
The stiff gas EOS, Eq. \eqref{eqn:stiff}, may be written in terms of energy as
\begin{equation}
e = \frac{p-c_s^2 (\rho - \rho_0)}{\rho(\gamma -1)}.
\end{equation}
\noindent Combining the EOS with the general solution, the shock speed is 
\begin{equation}
r_s(t) = \left[\frac{A_s +(3-\gamma ) \sqrt{\rho_0} u_0}{4 \sqrt{\rho_0}} \right]t 
\end{equation}
\noindent where 
\begin{equation}
A_s = \sqrt{16 \left(c^2 \rho_0+\gamma  p_0\right)+(\gamma +1)^2 \rho_0 u_0^2}.
\end{equation}
\noindent The density in the unshocked region is $\rho_1 = \rho_0$ and the density in the shocked region is
\begin{equation}
\rho_2 = \rho_0 \left[1-\frac{4 \sqrt{\rho_0} u_0}{A_s+(3-\gamma )  \sqrt{\rho_0} u_0}\right].
\end{equation}

\subsection{Noble-Abel Gas Solution}
The Noble-Abel EOS is Eq. \eqref{eqn:NobleAbel} may be written in terms of energy as
\begin{equation}
e = \frac{p(1-b \rho)}{\rho (\gamma -1)}.
\end{equation}
Using the general Noh solution and the EOS, the shock speed is 
\begin{equation}
r_s(t) = \left[ \frac{2 \rho_0 u_0^2 \left(2 b \rho_0+\gamma -1 \right)+4 \gamma  p_0}{A_{NA}+\rho_0 u_0 \left(4 b \rho_0+\gamma -3 \right)} \right] t
\end{equation}
where 
\begin{equation}
A_{NA} = \sqrt{\rho_0 \left[ \left(\gamma +1 \right)^2 \rho_0 u_0^2-16 \gamma  p_0 \left(b \rho_0-1 \right) \right]}.
\end{equation}
\noindent The density for the planar solution for the unshocked region is $\rho_1 = \rho_0$
and the density for the planar solution for the shocked region is
\begin{equation}
\rho_2 = \frac{\rho_0 \left[-u_0 A_{NA} +4 \gamma  p_0+\left(\gamma +1\right) \rho_0 u_0^2\right]}{2 \rho_0 u_0^2 (2 b \rho_0+\gamma   -1)+4 \gamma  p_0}.
\end{equation} 

As with the ideal gas solution, for $m=1$ or $2$ Noble-Abel gas solutions exists only in the infinitely strong shock case ($K_0 = 0$, $p_0 =0$) \cite{ramsey}. The shock speed is

\begin{equation}
r_s(t) = \frac{\left(\gamma -1 \right) u_0 t}{2 \left( b \rho_2 -1 \right)}.
\end{equation}

\noindent The density for the Noble-Abel gas in the unshocked region is
\begin{equation}
\rho_1 =  \rho_0 \left( 1- \frac{u_0 t}{r} \right)^m
\end{equation}

\noindent and the density for the Noble-Abel gas in the shocked region is
\begin{equation}
\rho_2 = \rho_0 \left( \frac{\gamma +1-2b \rho_2}{\gamma -1}\right)^{m+1}.
\end{equation}

\subsection{Carnahan-Starling Gas Solution} \label{CS}
The Carnahan-Starling gas EOS, Eq. \eqref{eqn:CarnStar}, may be rearranged in terms of energy as
\begin{equation}
e = \frac{p}{\rho(\gamma - 1)Z}.
\end{equation}
Combining this with Eqs. \eqref{eqn:unshocked}-\eqref{eqn:shockspeed2} the planar Carnahan-Starling fluid solution for the unshocked region is
\begin{align}
\rho_1 &= \rho_0 \\
p_1 &= p_0 \\
e_1 &= \frac{p_0}{\rho_0(\gamma - 1)Z(\rho_0)} \\
u_1 &= u_0.
\end{align}
Given $u_2 = 0$, the variables, $a$, $\rho_2$, $p_2$ and $e_2$ can be solved implicitly from the system, 
\begin{align}
\rho_2 &= \rho_0 \left(1-a u_0  \right)  \\
p_2 &= p_0 - \frac{\rho_0 u_0 }{a} \left(1-a u_0 \right)\\
e_2 &= e_1 + \frac{1}{2}u_0^2 - \frac{au_0p_0}{\rho_2} \\
e_2 &= \frac{p_2}{\rho_2(\gamma - 1)Z(\rho_2)}.
\end{align} 
For the convergence studies on the Carnahan-Starling EOS, we employed the Powell Hybrid Method using the fsolve function via the optimize package in Python \cite{fsolve}. The accuracy of the implicit solver ($\sim 10^{-8}$) makes the numerical solutions viable as quasi-analytic solutions for verification studies. 

\section{Verification}
Verification studies compare numerical implementations with analytically derived solutions. These studies are done in order to evaluate the accuracy and performance of numerical techniques and algorithms, in this case, for a hydrocode based on conservation laws. The FLAG code has been subject to rigorous verification and validation research. Some of the more recent efforts include the studies of the invariance with respect to the heat conduction in the Coggeshall problem \cite{hendon}, the Guderley shock problem \cite{ramsey2017}, self-similar shock wave propagation \cite{ramsey2016}, the impact of artificial viscosity on the Sedov problem \cite{doebling}, the Sedov shock wave blast \cite{pederson}, and the motivation behind the Godunov-like staggered grid hydrodynamic approach used in FLAG \cite{morgan}. The common theme of all these investigations (which is exemplary of the larger body of work performed on the code to date) centers around the use of the ideal gas EOS coupled to the code's hydrodynamic solver. Thus the motivation of our work: examples of verification studies performed using the same code and compressible-flow solver, but coupled to simple non-ideal EOS classes.

\subsection{FLAG Hydrocode} The Los Alamos Lagrangian hydrocode FLAG uses a compressible-flow algorithm featuring artificial viscosity and ancillary grid stability methods. The solver implements explicit, geometry-compatible, finite volume spatial full-Lagrangian staggered-grid hydrodynamics. This method uses control volumes with zones that are assigned by drawing lines between nodes. The densities, pressures, and specific internal energies live at zone centers and the velocities live at the nodes. Time integration features a predictor-corrector method based on pressure or energy projection for momentum evolution, as discussed by Burton \cite{burton90} and Caramana et al. \cite{caramana}. For smooth (i.e., shock-free) gas dynamics problems, the hydrodynamics solver is second-order accurate in space and time (further assuming a zero linear artificial viscosity coefficient) \cite{burton90, burton92, burton94, caramana}. 

FLAG is a high-resolution numerical scheme with accuracy limited by the presence of the shock, which by Godunov's order barrier theorem restricts the studies to first-order spatial convergence \cite{godunov54, godunov59}.  Our hydrocode studies use the Barton artificial viscosity model. In compression, the quadratic viscosity coefficient is 2.0 and the linear viscosity coefficient is 0.3. The linear viscosity coefficient in expansion is 0.3 \cite{vonneumann1950method, richtmyer1967k}. (In the context of the Noh problem, the FLAG algorithms become equivalent to Reimann-solver based methods \cite{menikoff1989riemann, banks2010exact, whitham2011linear, toro2013riemann}). These constructs are used to capture the zero-thickness of the jump discontinuity on a grid of finite-thickness and the nonzero linear coeffiencient are also responsible for expectation that the spatial convergence is first order \cite{majda, banks}.  Another factor that alters the formation of the shock profile is the wall heating produced by the shock reflecting at the origin.  This is caused by the pressure initially being computed and then energy/temperature added to the system, via the EOS, to compensate for a shortage in density \cite{rider}.
The convergence order will remain first order despite refining the spatial mesh size because of the presence of the artificial viscosity and Gudunov's theorem.   \\

\subsection{Verification Studies}

\begin{figure}[H]
\centering  
\subfigure[$p_0=\SI{0.00}{Mbar}$]{\includegraphics[width=2.5in]{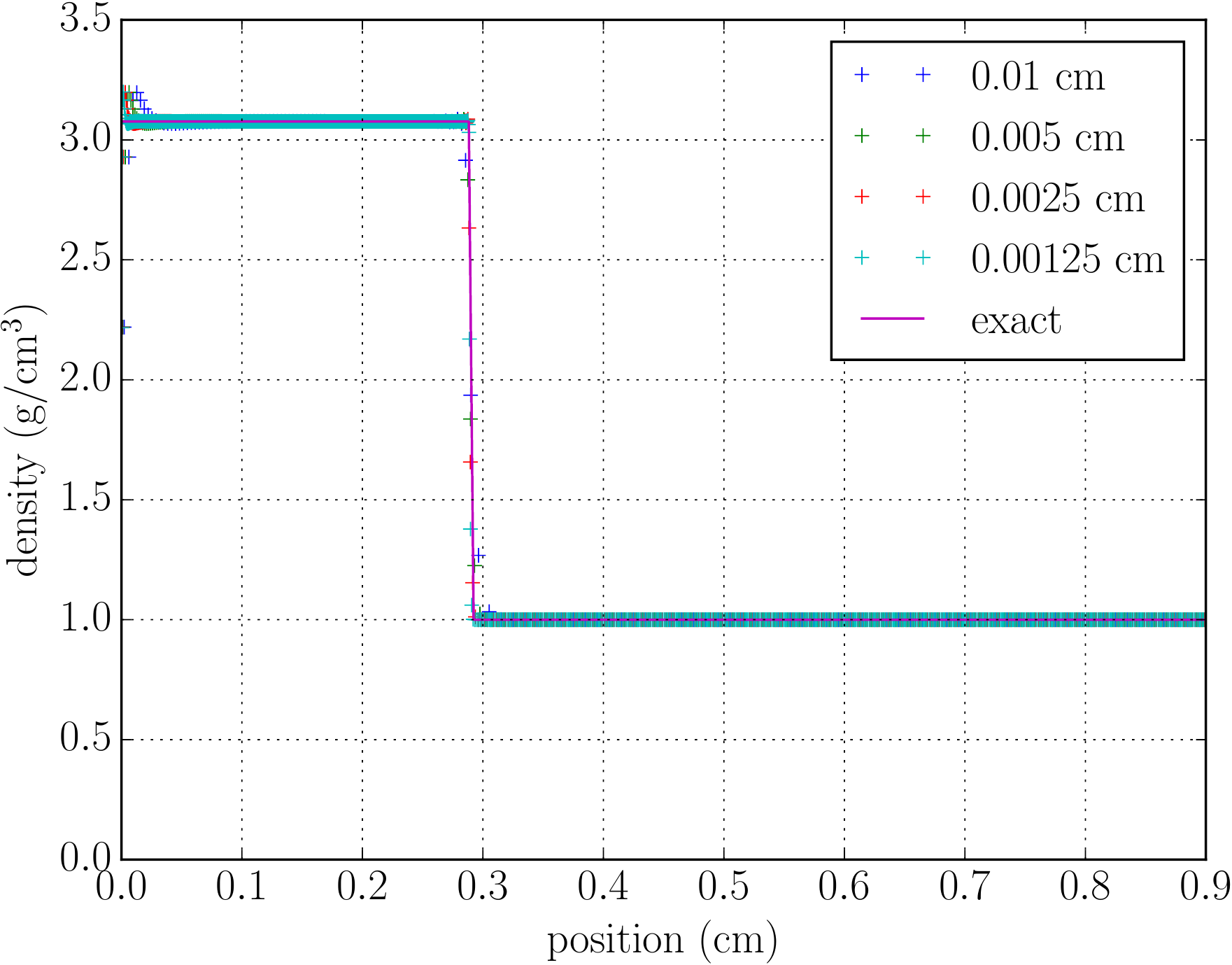}} 
\subfigure[$p_0=\SI{0.01}{Mbar}$]{\includegraphics[width=2.5in]{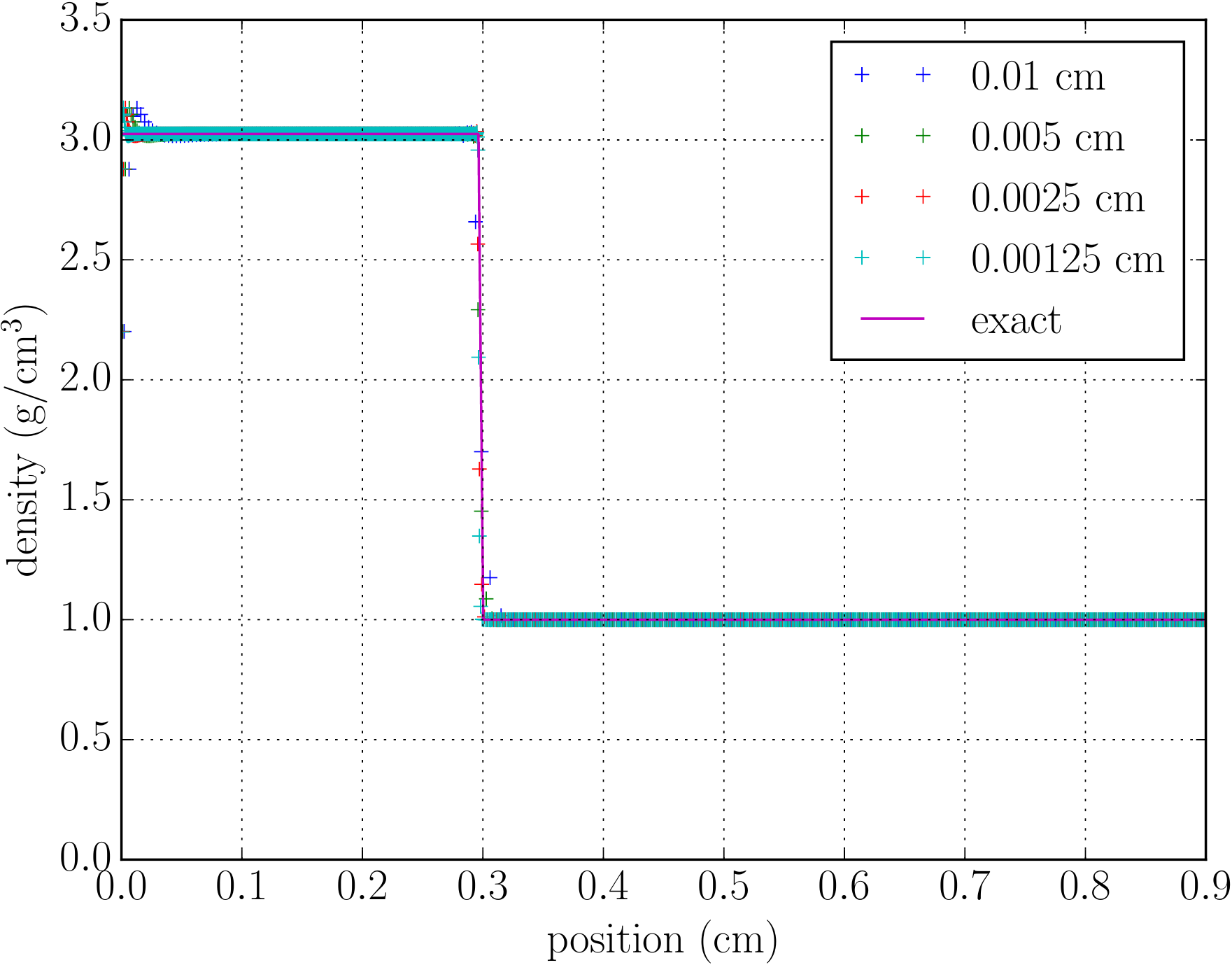}}
\subfigure[$p_0=\SI{0.10}{Mbar}$]{\includegraphics[width=2.5in]{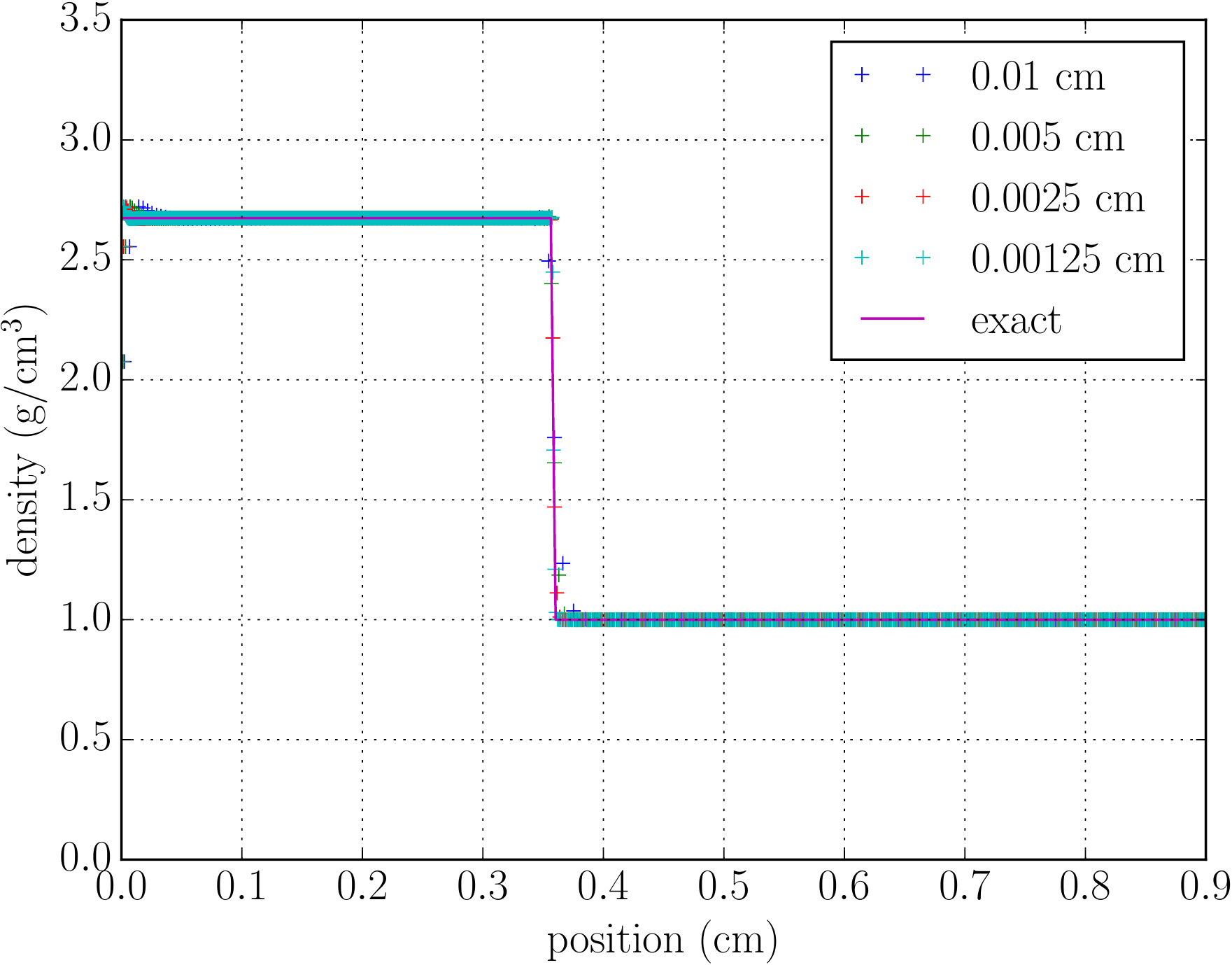}}
\subfigure[$p_0=\SI{1.00}{Mbar}$]{\includegraphics[width=2.5in]{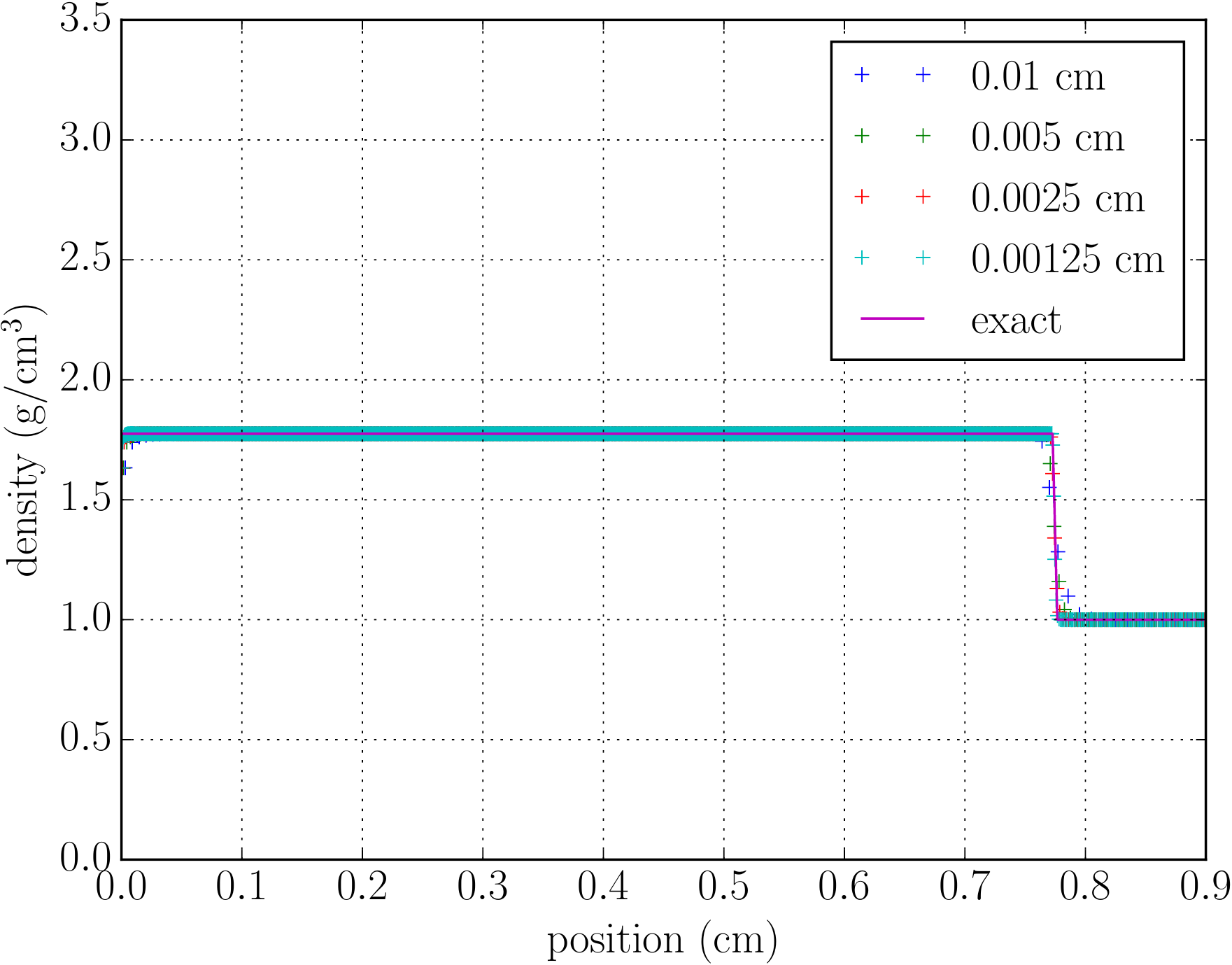}}
\caption{Density plots of planar Noble-Abel gas, $\gamma = 5/3$, $\rho_0 = \SI{1.0}{g/cm^3}$, $b=\SI{0.1}{cm^3/g}$}
\label{fig:NApvary}
\end{figure}

\begin{figure}[H]
\centering
\subfigure[$b=\SI{-0.2}{cm^3/g}$]{\includegraphics[width=2.5in]{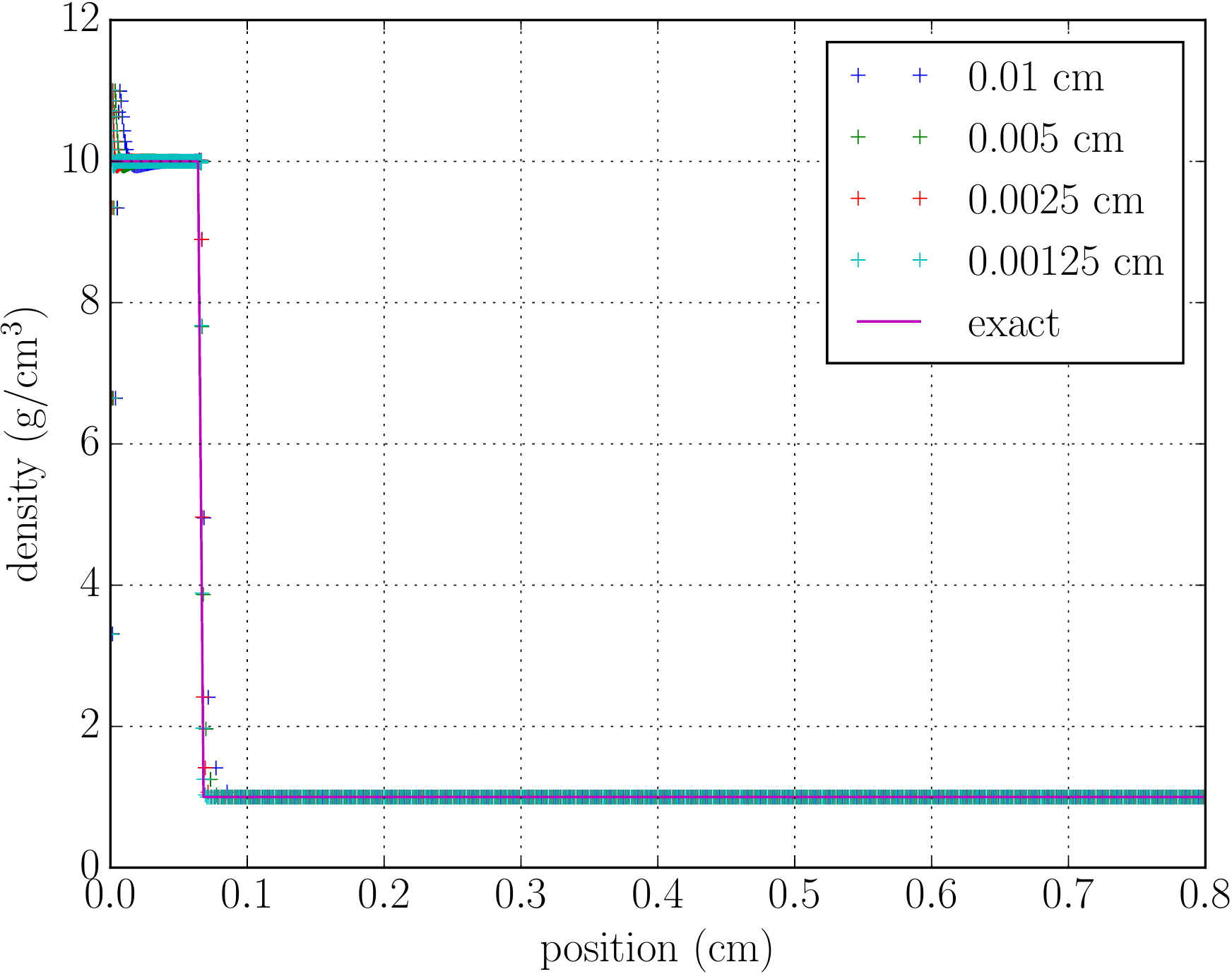}} 
\subfigure[$b=\SI{0.0}{cm^3/g}$]{\includegraphics[width=2.5in]{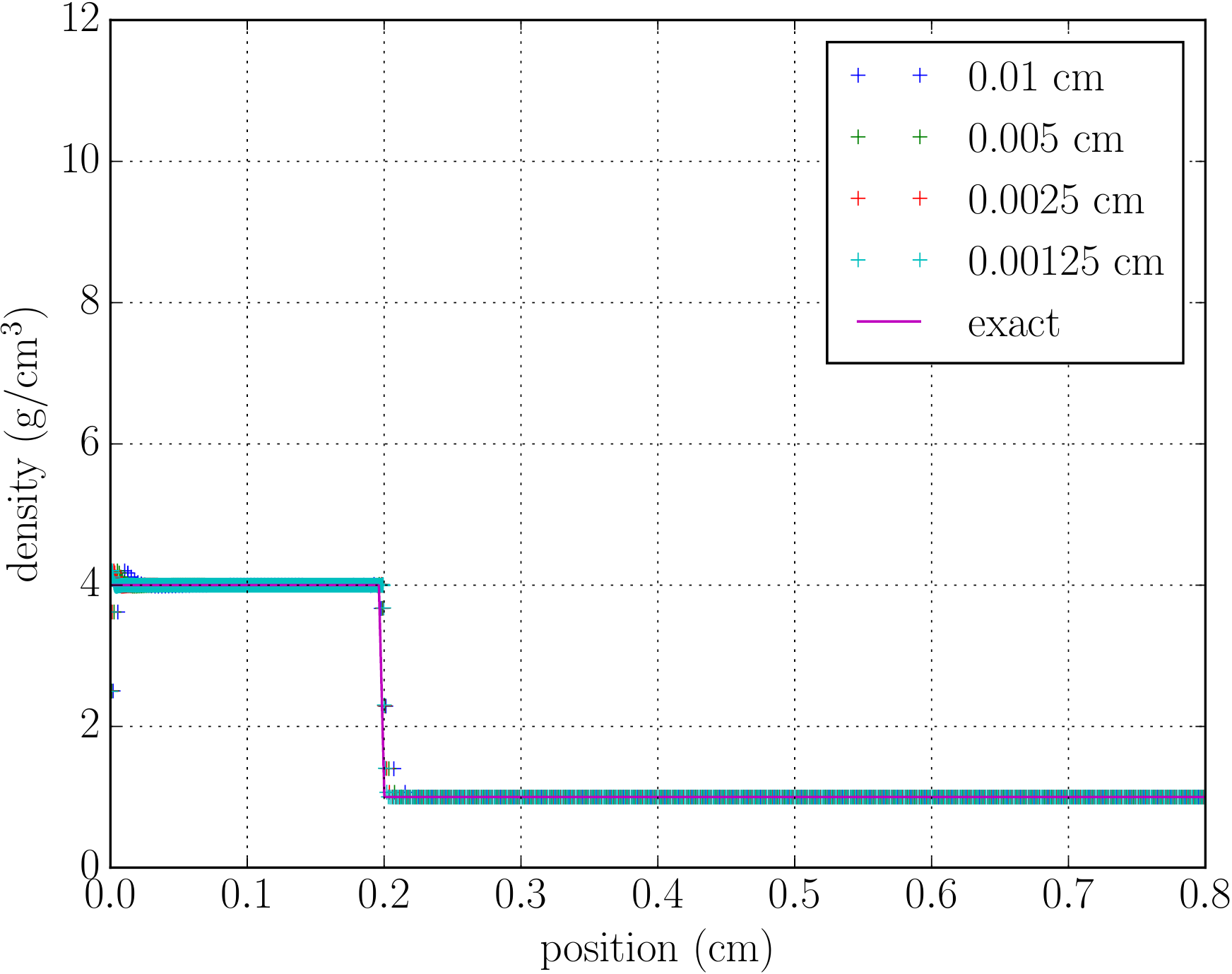}}
\subfigure[$b=\SI{0.2}{cm^3/g}$]{\includegraphics[width=2.5in]{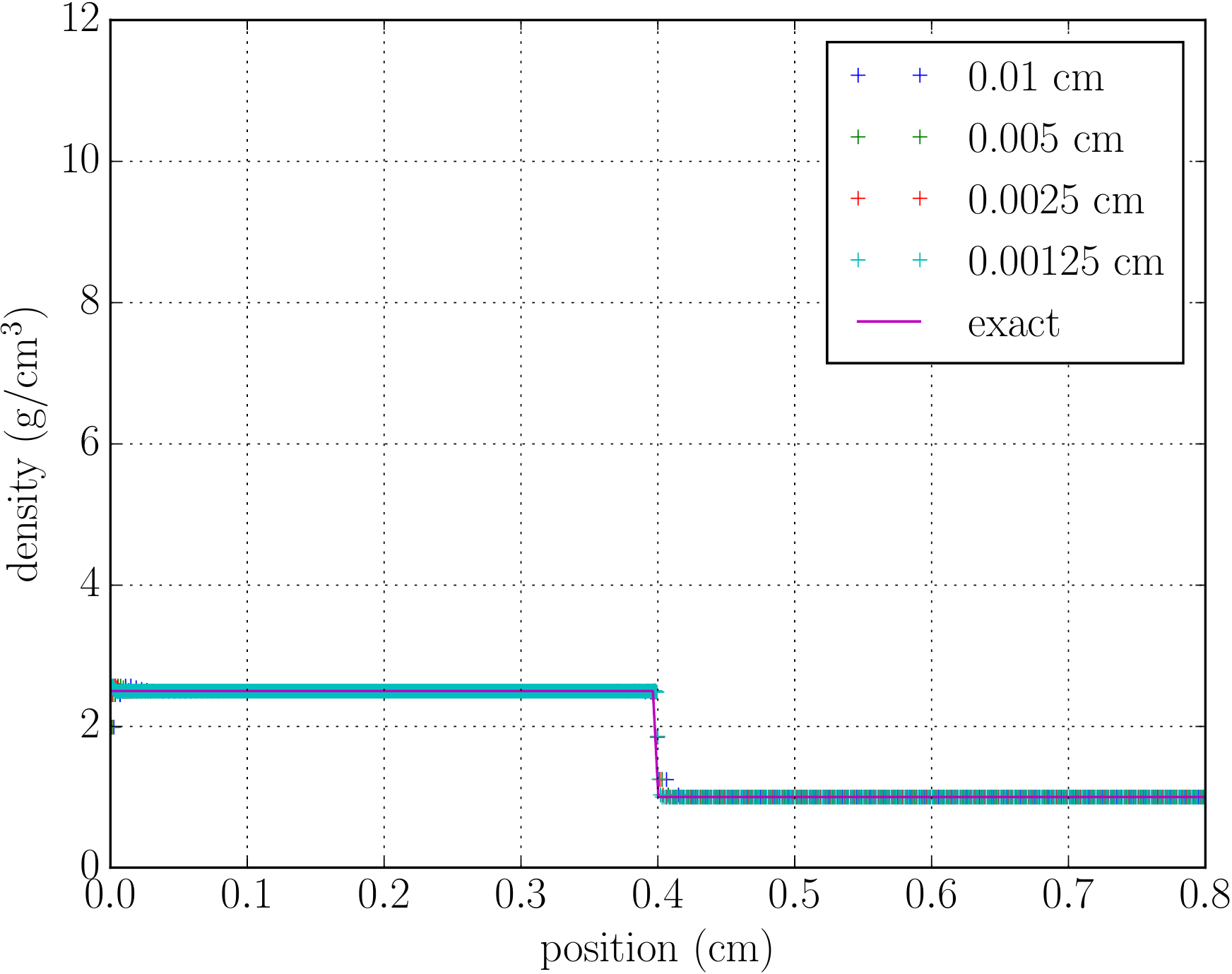}}
\subfigure[$b=\SI{0.4}{cm^3/g}$]{\includegraphics[width=2.5in]{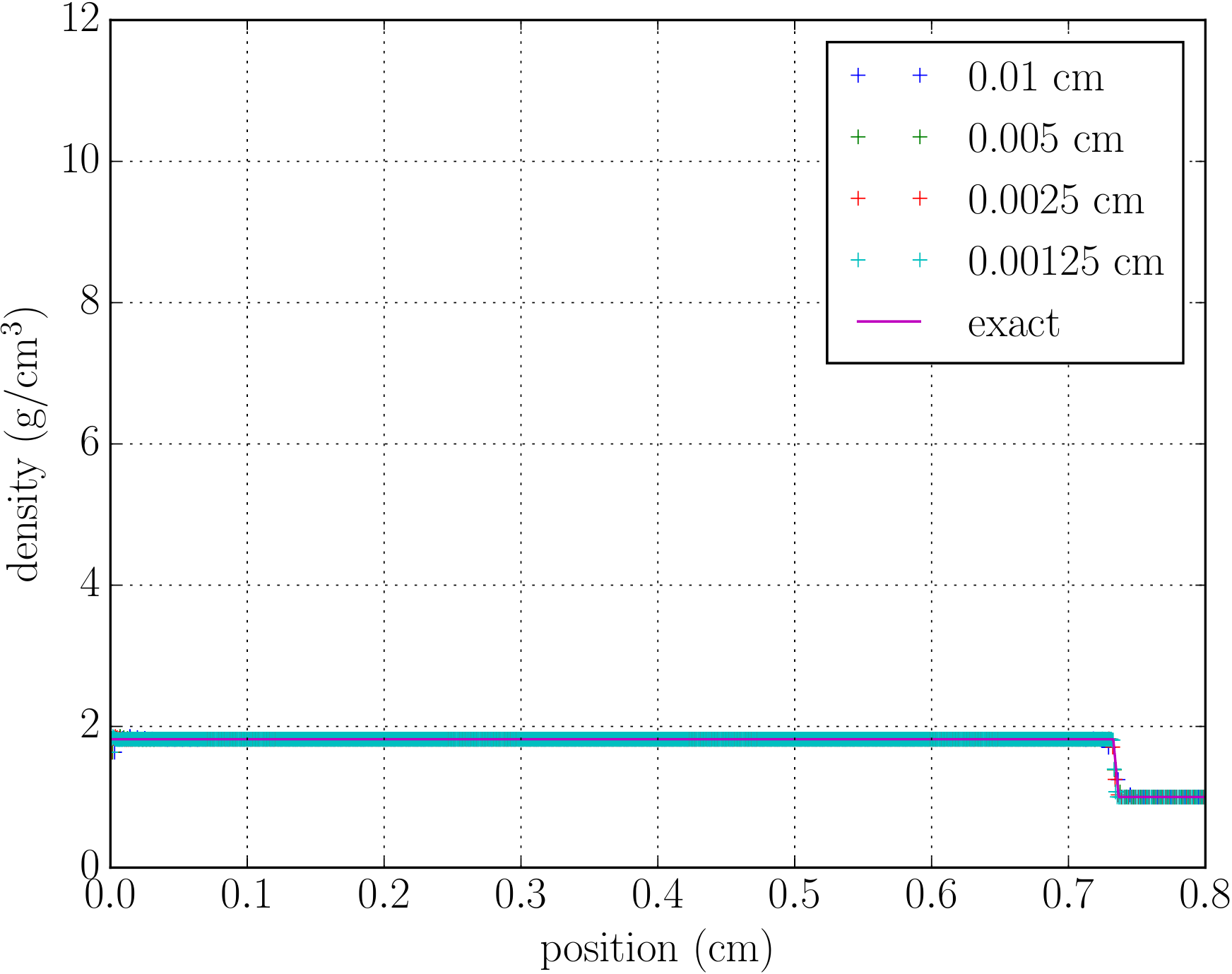}}
\caption{Density plots of planar Noble-Abel gas, $\gamma = 5/3$, $\rho_0 = \SI{1.0}{g/cm^3}$, $p_0=\SI{0.0}{Mbar}$}
\label{fig:NAbvary}
\end{figure}

\begin{figure}[H]
\centering  
\subfigure[$b=\SI{0.00}{cm^3/g}$]{\includegraphics[width=2.5in]{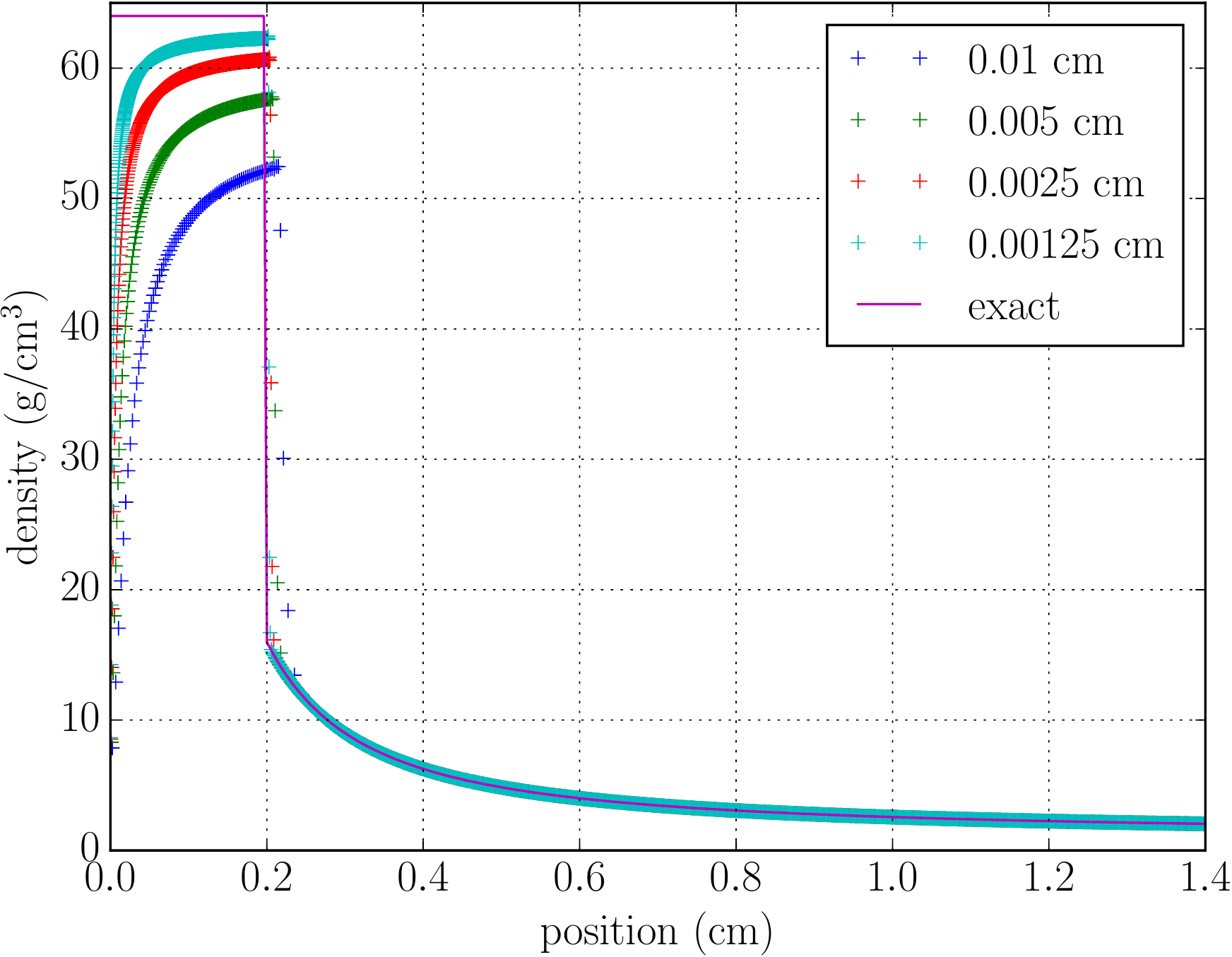}} 
\subfigure[$b=\SI{0.01}{cm^3/g}$]{\includegraphics[width=2.5in]{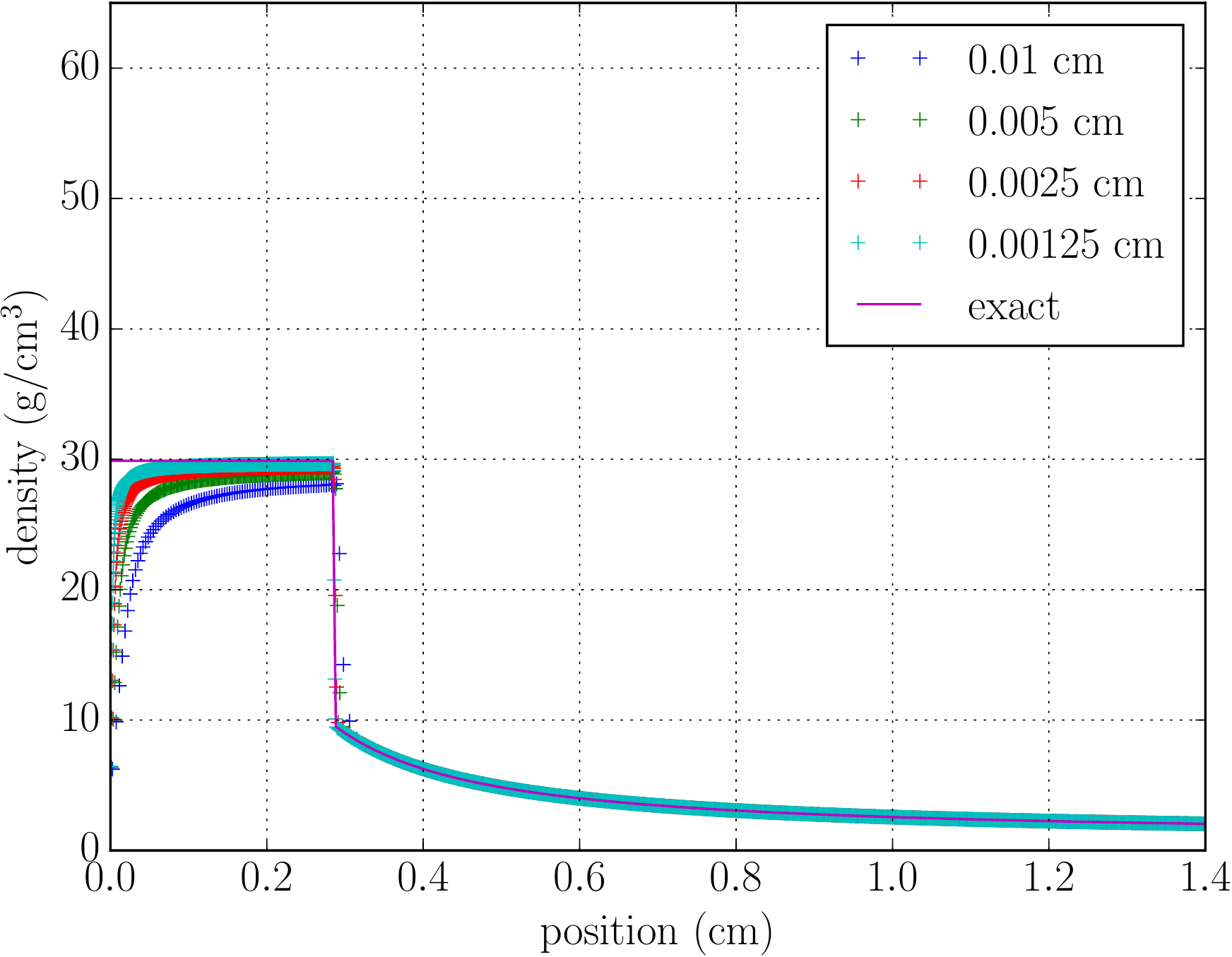}}
\subfigure[$b=\SI{0.10}{cm^3/g}$]{\includegraphics[width=2.5in]{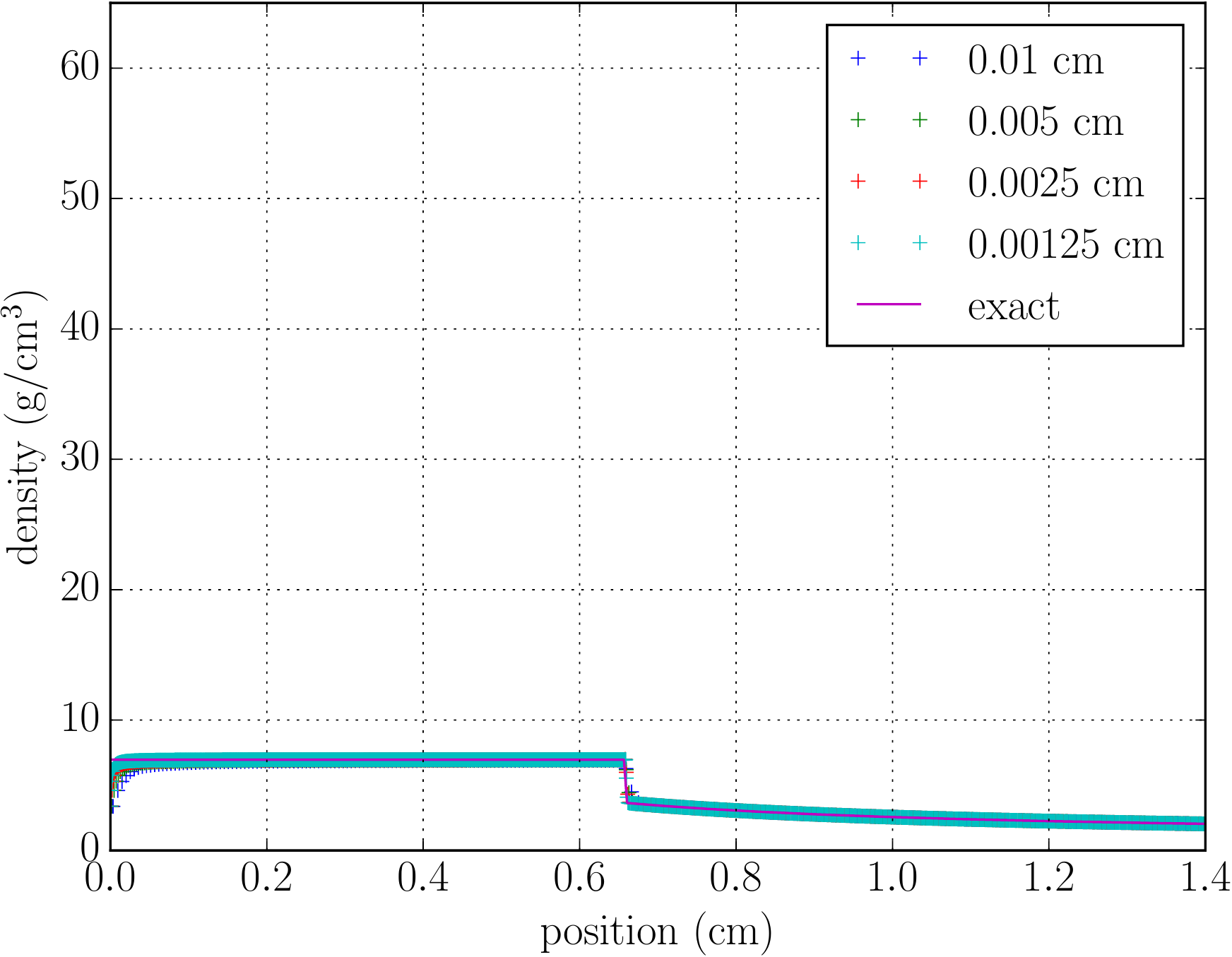}}
\subfigure[$b=\SI{0.20}{cm^3/g}$]{\includegraphics[width=2.5in]{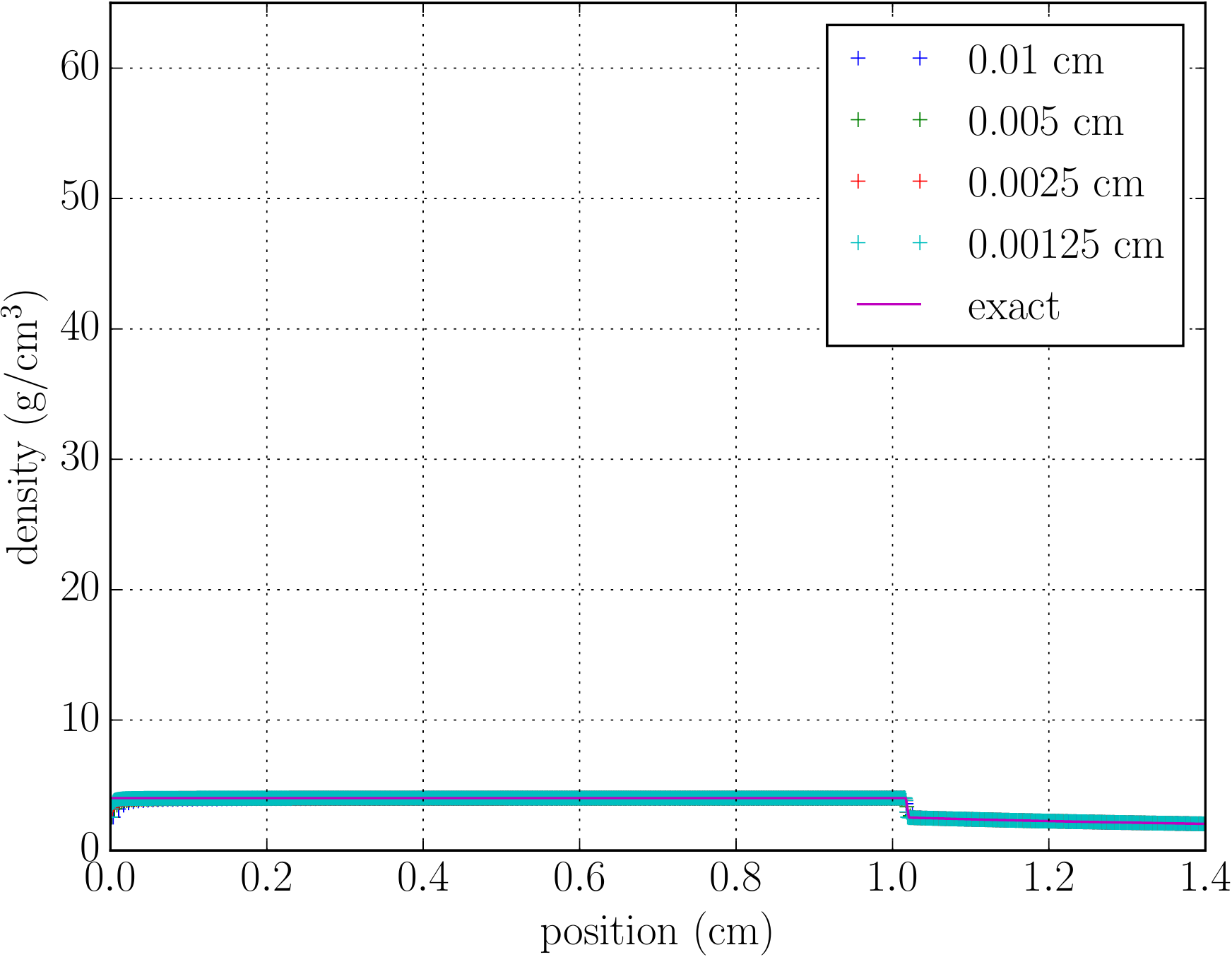}}
\caption{Density plots of spherical Noble-Abel gas, $\gamma = 5/3$, $\rho_0 = \SI{1.0}{g/cm^3}$, $p_0=\SI{0.0}{Mbar}$}
\label{fig: NA3D}
\end{figure}

\begin{figure}[H]
\centering 
\subfigure[$b=\SI{0.00}{cm^3/g}$]{\includegraphics[width=2.5in]{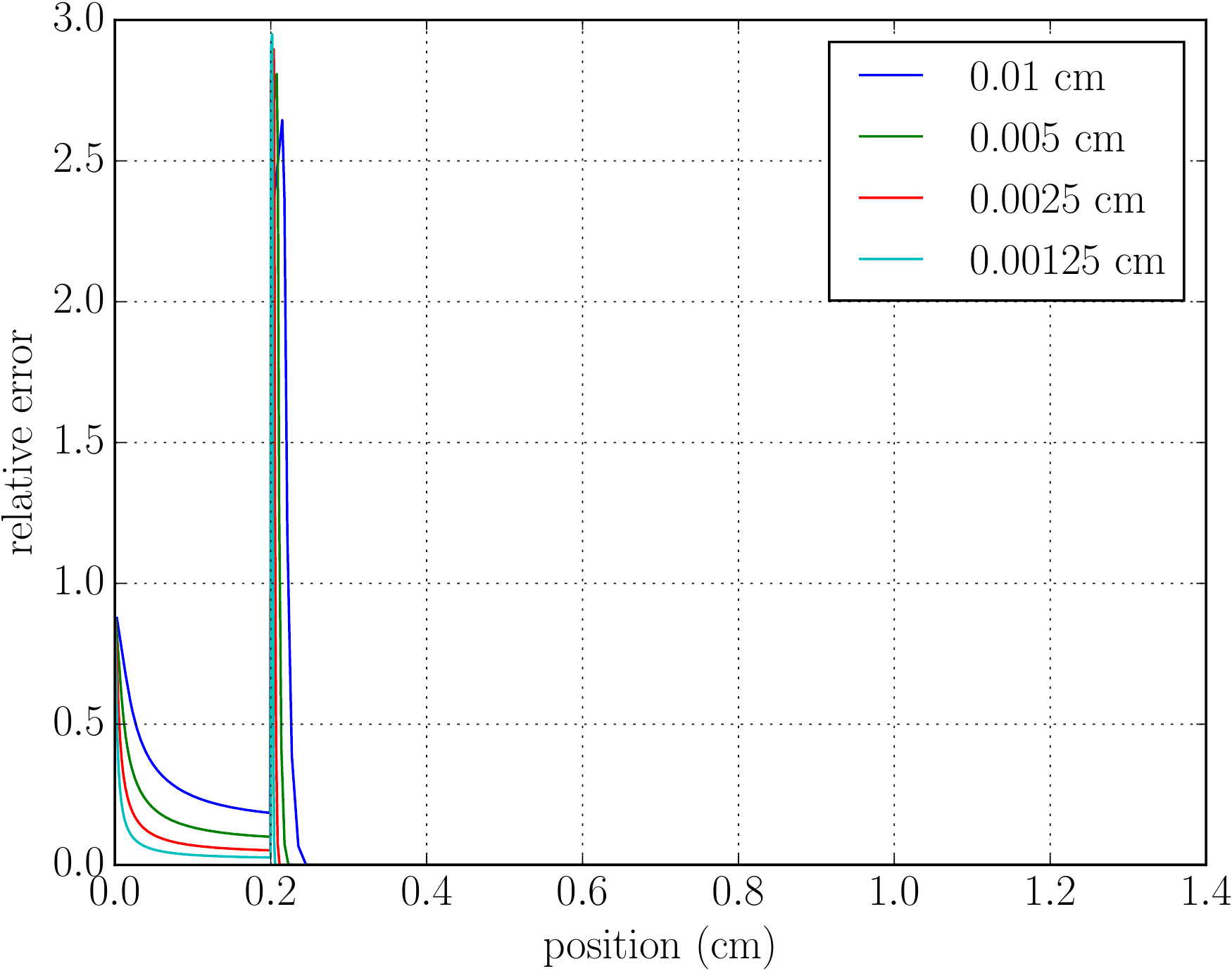}} 
\subfigure[$b=\SI{0.01}{cm^3/g}$]{\includegraphics[width=2.5in]{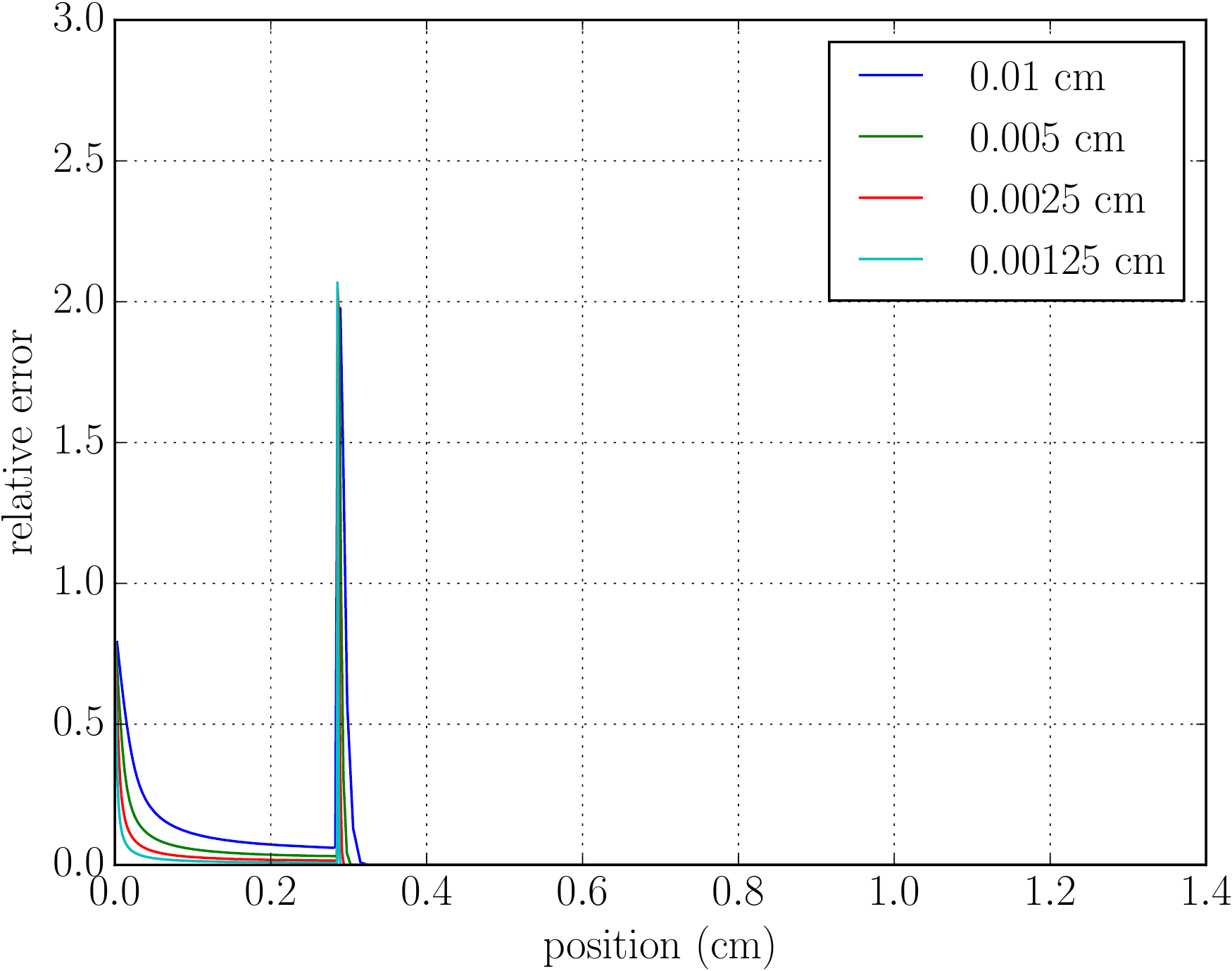}}
\subfigure[$b=\SI{0.10}{cm^3/g}$]{\includegraphics[width=2.5in]{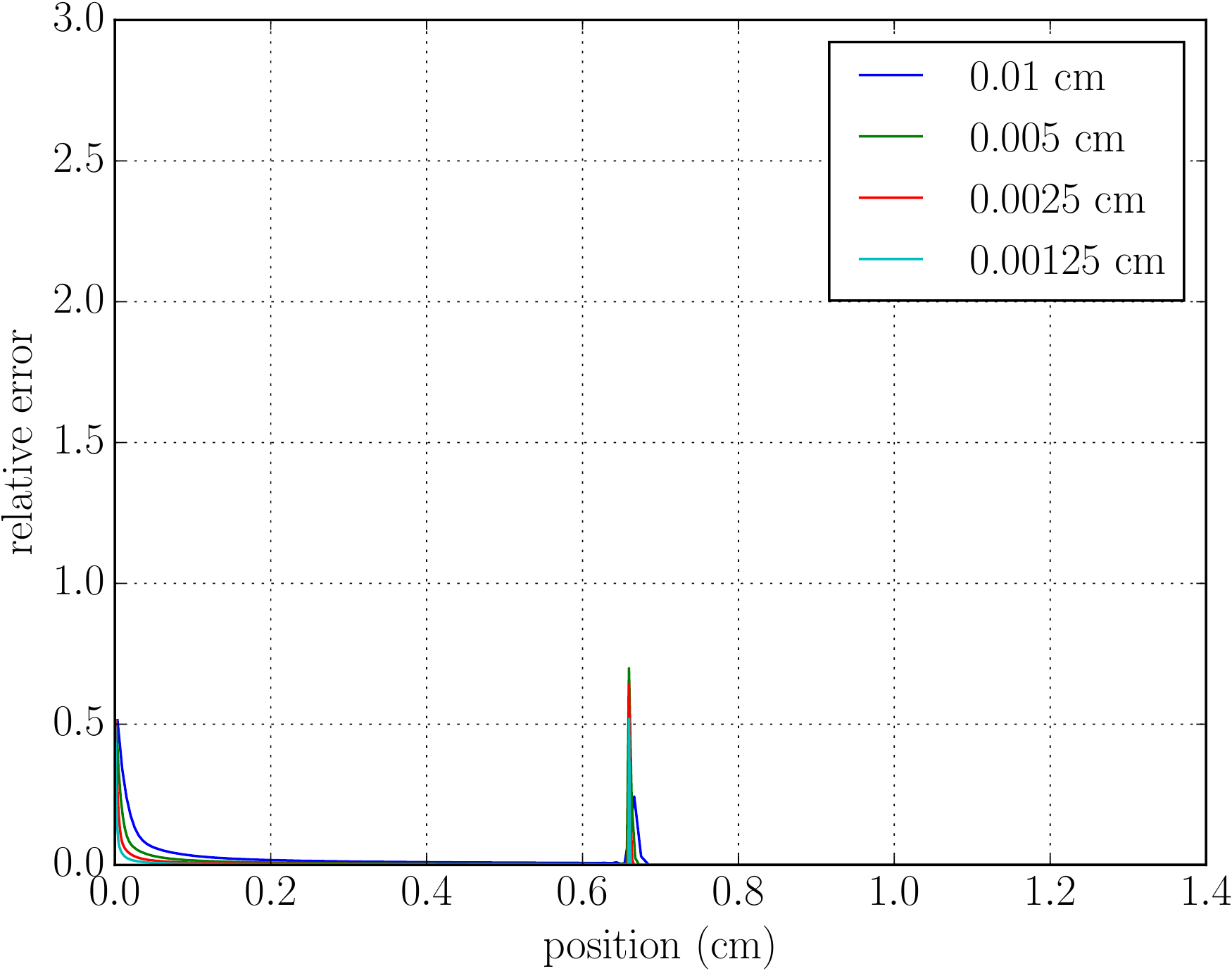}}
\subfigure[$b=\SI{0.20}{cm^3/g}$]{\includegraphics[width=2.5in]{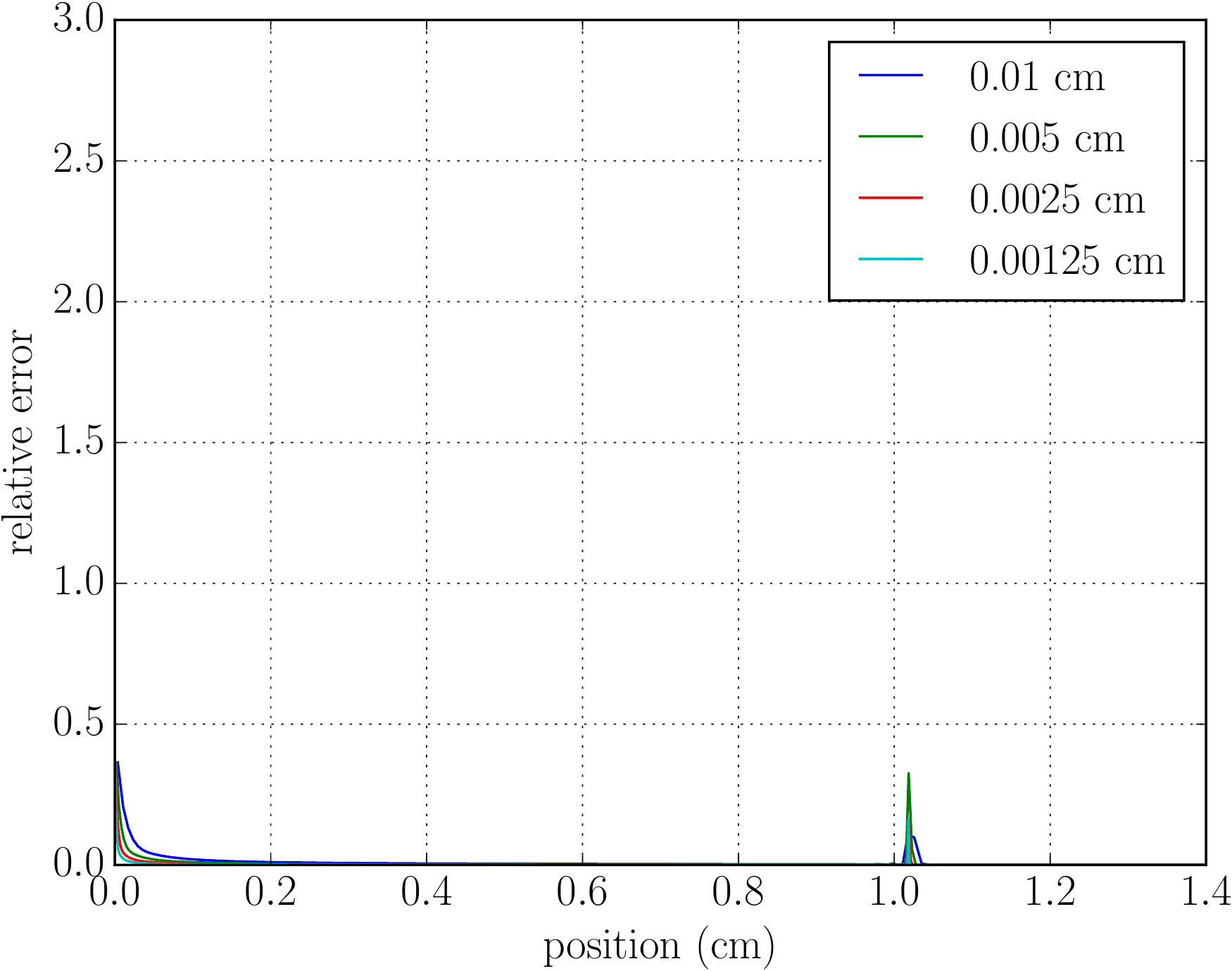}}
\caption{Relative error plots of spherical Noble-Abel gas, $\gamma = 5/3$, $\rho_0 = \SI{1.0}{g/cm^3}$, $p_0=\SI{0.0}{Mbar}$}
\label{fig: NA3D_err}
\end{figure}

\begin{figure}[H]
\centering
\includegraphics[width=3.25in]{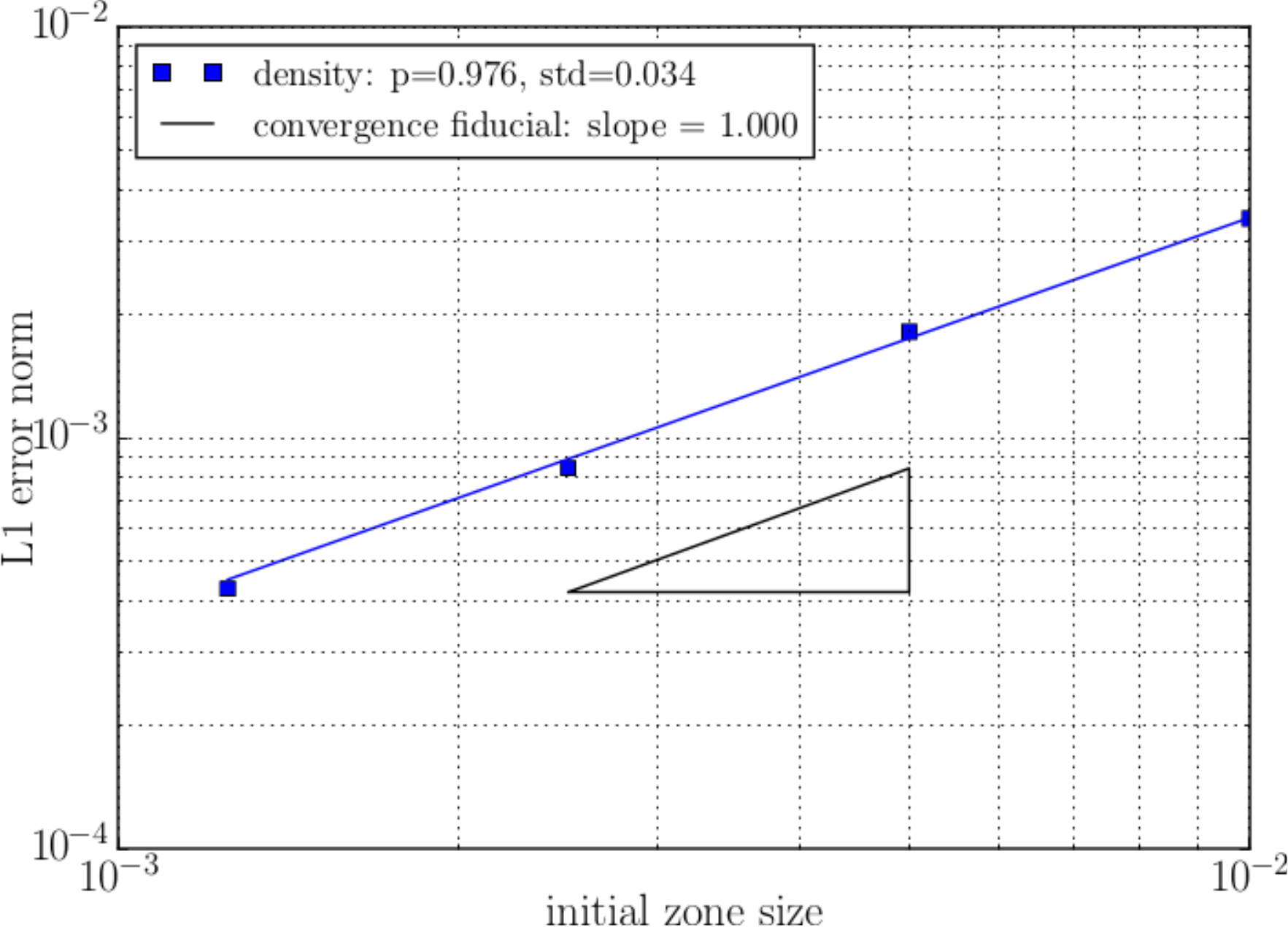}
\caption{Convergence study for the planar Noble-Abel case when $p_0=\SI{1}{Mbar}$ and $b=\SI{0.4}{cm^3/g}$.} \label{fig:convergencestudy}
\end{figure}

FLAG results for the pressure, density, and specific internal energy as a function of time, position, and mesh resolution are compared with the exact solutions for the ideal and the non-ideal gases.  
Representative results for the density as a function of time are shown in Figs. \ref{fig:NApvary}-\ref{fig:convergencestudy} for the Noble-Abel fluid as a function of initial pressure, co-volume, and mesh size. The plots were produced using the verification tool ExactPack \cite{singleton, ExactPack}.
All results use the same initial density $\rho_0 = \SI{1}{g/cm^3}$, the same adiabatic coefficient $\gamma = 5/3$ and are measured at time $t=\SI{0.6}{\mu s}$. 

In \ref{fig:NApvary}-\ref{fig: NA3D_err} the points are the FLAG results at indicated mesh resolutions ($\Delta r = 0.01$, $0.005$, $0.0025$, and $\SI{0.00125}{cm}$) and the lines represents the exact solution. In Figs. \ref{fig:NApvary}-\ref{fig: NA3D_err}, results are shown for the Noble-Abel EOS in the planar and spherical cases. Figure \ref{fig: NA3D_err} is the corresponding relative error between the FLAG results and the exact solution corresponding to the studies in Fig. \ref{fig: NA3D}.  For the cylindrical and spherical cases, note the simulation remains a one-dimensional with modifications for the converging geometry taken into consideration. 

In every case, the code behaves consistently with the exact solution except at wall and the shock front -- deficiencies common to many hydrocodes. The representative plots shown here for the Noble-Abel EOS display the same trends as the other EOSs. The simplest trend is the effect of the increase in initial pressure on the shock speed and density. In all cases, as the unshocked pressure increases, the shocked density decreases and the shock speed increases. Increasing the initial pressure increases the sound speed at all densities, which then leads the shock to move faster. Also, as the co-volume, $b$, increases in the Carnahan-Starling gas and Noble-Abel gas (Fig. \ref{fig:NAbvary}, \ref{fig: NA3D}) or as the reference sound speed, $c_s$, increases in the material of the stiff gas, the density in the shocked region decreases and the shock speed increases. When the reference sound speed of the stiff gas increases, the shock moves faster for the same reason as when the initial pressure is increased. By increasing the co-volume in the Noble-Abel gas and the Carnahan-Starling gas, the fluid becomes less compressible resulting in a smaller density jump across the shock front and faster moving wave.  
Conversely, when these EOSs have a negative co-volume, the molecules attract each other resulting in higher shocked densities and lower shock speeds than the ideal gas.  \\

The simulation results also reveal additional correlations between the various EOS parameters and the most prominent computed solution errors. From the relative error plots shown in Fig. \ref{fig: NA3D_err}, the computed errors in the density at $r=0$ and at the shock front decrease as the co-volume, $b$, increases. Similar results were found when increasing the EOS parameter co-volume and initial pressure for the other EOSs. At first glance, it might appear that the drop in relative error at the discontinuity is connected to the shock's position relative to the origin; the closer the shock is to the origin, the more wall heating influences the precise shape of the shock front. Upon closer investigation, by looking at a shock at earlier and later times, we found that the distribution of the relative error does not change. Therefore, we can conclude that the compressibility or sound speed in the fluid is what changes the distribution of the relative error at the shock front, not simply the distance from the wall. Higher co-volumes, sound speeds, and initial pressures all are observed to reduce the magnitude of the error at the wall and the jump discontinuity. \\

For infinitely strong shock cases, shown for the Noble-Abel gas in Figs. \ref{fig: NA3D_err}, the wall heating increases with increasing curvilinearity (planar to spherical). 
In cylindrical and spherical geometries,  the change in the compression term, $b$, more dramatically affects the wall heating than it does in the planar case due to the additional dissipation associated with the flow converging on an axis or a single point. \\

\twocolumn
\clearpage

First-order spatial convergence with respect to mesh size is observed with $\Delta r = 0.01$, $0.005$, $0.0025$, and $\SI{0.00125}{cm}$. The verification tool ExactPack \cite{singleton, ExactPack} analyzes the results by calculating the L1-norm \cite{oberkampf, roache}
\begin{equation}
\| y^E - y^F \| _1 \approx \frac{1}{N} \sum_{i=1}^Nw_i \left| y_i^E-y_i^F \right|
\end{equation}
where $y_E$ is the exact solution and $y_F$ is the FLAG data. The weights $w_i$ are determined by the dimensionality of the problem. For the 1D Cartesian problem, they are the zone lengths.  The L1-norm with respect to the mesh size is fitted on a log-log scale to a line in the form 
\begin{equation}
\ln \left(\| y^E - y^S \| _1 \right) =p \ln \left( \Delta x \right) +c. \label{eqn:slope}
\end{equation} 
Figure \ref{fig:convergencestudy} shows a typical result obtained from an ExactPack convergence order study for the Noble-Abel EOS.
The slope, $p$, is determined from Eq. \eqref{eqn:slope} and is known as the convergence order fitting this equation,
\begin{equation}
\| y^E - y^S \| _1 = e^c \left( \Delta x \right)^p.
\end{equation}
The convergence orders for all the EOS mentioned in this paper are presented in Tables 1-\ref{table: last}. Tables 1-2 and 4 are planar geometry while varying the EOS parameter and the initial pressure. Table \ref{table:NA} shows the results for the cuvilinear geometries ($m = 0, 1,$ and $2$) using the Noble-Abel EOS and $p_0 = 0$. \\

\begin{table}[H]
\centering
{\renewcommand{\arraystretch}{1.1}
\begin{tabular}{|c|| c c c c |} 
\hline
\backslashbox{$\mathbf{c_s}$}{$\mathbf{p_0}$} & \textbf{0.00} & \textbf{0.01} & \textbf{0.10} & \textbf{1.00} \\ \hline \hline
\textbf{0.0} & 0.998 & 0.987 & 1.017 & 0.997 \\
\textbf{0.1} & 0.993 & 1.062 & 0.989 & 0.974 \\
\textbf{0.5} & 0.998 & 0.978 & 1.001 & 0.988 \\
\textbf{0.6} & 1.018 & 0.988 & 1.017 & 0.987 \\
\textbf{1.0} & 1.011 & 0.975 & 1.008 & 1.039 \\
\textbf{1.5} & 0.973 & 1.087 & 0.975 & 0.989 \\
\textbf{2.0} & 0.987 & 1.019 & 1.019 & 0.985 \\
\hline
\end{tabular}}
\label{table: first}
\caption{Table of convergence orders for planar stiff gas. The units for $c_s$ are $cm/\mu s$ and for $p_0$ the units are $Mbar$. }
\vspace{2em}
{\renewcommand{\arraystretch}{1.1}
\begin{tabular}{|c|| c c c c |}
\hline
\backslashbox{\textbf{b}}{$\mathbf{p_0}$} & \textbf{0.00} & \textbf{0.01} & \textbf{0.10} & \textbf{1.00} \\ \hline \hline
\textbf{-0.20} & 1.031 & 1.083 & 1.028 & 0.955 \\
\textbf{-0.10} & 1.078 & 1.064 & 0.960 & 1.007 \\
\textbf{0.00} & 0.998 & 0.987 & 1.017 & 0.997 \\
\textbf{0.01} & 0.979 & 0.938 & 0.994 & 1.005 \\
\textbf{0.10} & 1.030 & 0.988 & 1.036 & 0.956 \\
\textbf{0.20} & 1.000 & 1.027 & 1.006 & 1.001 \\
\textbf{0.40} & 0.924 & 0.996 & 0.964 & 0.976 \\
\hline
\end{tabular}}
\caption{Table of convergence orders for planar Noble-Abel gas. The units for $b$ are $cm^3/g$. }
\vspace{2em}
{\renewcommand{\arraystretch}{1.1}
\begin{tabular}{|c|| c c c|}
\hline
\backslashbox{\textbf{b}}{\textbf{geometry}} & \textbf{planar} & \textbf{cylindrical} & \textbf{spherical} \\ \hline \hline
\textbf{0.000} & 0.998 & 0.970 & 0.900  \\
\textbf{0.001} & 1.003 & 0.960 & 0.925 \\
\textbf{0.010} & 0.979 & 0.985 & 1.045 \\
\textbf{0.100} & 1.030 & 0.858 & 0.771 \\
\textbf{0.200} & 1.000 & 0.865 & 0.907 \\
\textbf{0.400} & 0.924 & 1.253 & 0.923 \\
\hline
\end{tabular}}
\caption{Table of convergence orders for Noble-Abel gas for infinitely-strong shocks ($p_0 = 0$).}
\label{table:NA}
\vspace{2em}
{\renewcommand{\arraystretch}{1.1}
\begin{tabular}{|c|| c c c c |}
\hline
\backslashbox{\textbf{b}}{$\mathbf{p_0}$} & \textbf{0.00} & \textbf{0.01} & \textbf{0.10} & \textbf{1.00} \\ \hline \hline
\textbf{-0.10} & 0.954 & 1.139 & 0.963 & 0.994 \\
\textbf{0.00} & 0.998 & 0.987 & 1.017 & 0.997 \\
\textbf{0.01} & 1.061 & 1.008 & 1.070 & 1.015 \\
\textbf{0.05} & 0.903 & 0.984 & 1.044 & 0.999 \\
\textbf{0.10} & 0.978 & 1.049 & 1.050 & 0.963 \\
\textbf{0.20} & 0.961 & 1.000 & 0.913 & 0.993 \\
\textbf{0.30} & 1.166 & 1.040 & 1.045 & 0.973 \\
\textbf{0.40} & 0.900 & 0.919 & 0.884 & 0.976 \\
\hline
\end{tabular}}
\caption{Table of convergence orders for planar Carnahan-Starling gas. }
\label{table: last}
\end{table}

The results of Tables 1-4 fit our expectations of FLAG's performance. The convergence studies of the code in the majority cases were close to being spatially first-order accurate.

\section{Conclusion}

Our results illustrate how non-ideal EOSs can provide a more physically-realistic  complement to the ideal gas in verifying complex hydrocodes. Proper verification studies that feature shocks typically exhibit first-order spatial convergence \cite{majda, banks}. 
Our studies demonstrate that the LANL hydrocode FLAG models the Noh solution with first-order accuracy for both ideal and several non-ideal fluids when compared to the exact solutions. 
Additionally, the decrease in the amount of error as the shock speed increases is significant. We observe that the higher the initial pressure and/or the larger the initial bulk modulus, the smaller the magnitude of the error both at the wall and the shock front.  \\

Further analysis could be done by examining pointwise verification metrics such as the shock location as a function of the non-ideal parameters, examining other EOSs, varying the adiabatic constant, examining shorter time scales, and varying the initial density.  
In addition to modifying parameters, it would be interesting to compare to a verification study utilizing an Eulerian hydrocode such as the LANL code, RAGE \cite{gittings2008rage}. Research has already been published on the stiff gas EOS using the RAGE hydrocode\cite{yorke}. The Eulerian code was shown to give better results for the Noh problem than the Lagrangian code as discussed by Rider \cite{rider}. In our results, the lowest convergence orders were seen in the curvilinear cases so this should be the focus of the comparison study. Additionally, one might undertake multidimensional studies and see how the results compare to the 1D curvilinear cases. Finally, exact solutions can be derived by applying Lie group methods to other problems such as the Noh-2 problem \cite{noh}, the Sedov problem \cite{kamm, hutchens}, the Guderley problem \cite{guderley,schmidt,ramsey2017,holm}, and the Sod Problem \cite{sod}. Verification studies would be needed to show that the FLAG hydrocode performed as expected for these problems since they are considered more complex than the Noh problem.


\begin{acknowledgment}

This work was performed under the auspices of the United States Department of Energy by Los Alamos National Security, LLC, at Los Alamos National Laboratory under Contract No. DE-AC52-06NA25396. Extensive use of the LANL verification tool ExactPack was made in this study.  The authors gratefully acknowledge the support of the U.S. DOE Advanced Strategic Computing Program, in particular the Physics Verification Project and Physics and Engineering Models Project. \\

This material is based upon work supported by the National Science Foundation Graduate Research Fellowship Program under Grant No. DGE 1322106. Any opinions, findings, and conclusions or recommendations expressed in this material are those of the author(s) and do not necessarily reflect the views of the National Science Foundation.

\end{acknowledgment}

\bibliographystyle{asmems4}

\bibliography{Bibliography.bib}


\end{document}